\documentclass[prb,aps,amsmath,amssymb,showpacs,twocolumn]{revtex4}

\usepackage{bm,color,float,dcolumn,amsmath,amssymb,amsfonts,amsthm,graphicx,hyperref,subfigure,dsfont}

\def\be{\begin{equation}}
\def\ee{\end{equation}}
\def\bea{\begin{eqnarray}}
\def\eea{\end{eqnarray}}

\begin{document}

\title{Theoretical investigation of edge reconstruction in the $\nu=\frac{5}{2}$ and $\frac{7}{3}$ fractional quantum Hall states}

\author{Yuhe Zhang, Ying-Hai Wu, Jimmy A. Hutasoit, and Jainendra K. Jain}

\affiliation{Department of Physics, The Pennsylvania State University, University Park, PA 16802, USA}

\date{\today}

\begin{abstract}
The edge physics of the $\nu=5/2$ fractional quantum Hall state is of relevance to several recent experiments that use it as a probe to gain insight into the nature of the bulk state. We perform calculations in a semi-realistic setup with positive background charge at a distance $d$, by exact diagonalization both in the full Hilbert space (neglecting Landau level mixing) and in the restricted Pfaffian basis of edge excitations. Our principal finding is that the $5/2$ edge is unstable to a reconstruction except for very small $d$. In addition, the interactions between the electrons in the second Landau level and the lowest Landau level enhance the tendency toward edge reconstruction. We identify the bosonic and fermionic modes of edge excitations and obtain their dispersions by back-calculating from the energy spectra as well as directly from appropriate trial wave functions. We find that the edge reconstruction is driven by an instability in the fermionic sector for setback distances close to the critical ones. We also study the edge of the $\nu=7/3$ state and find that edge reconstruction occurs here more readily than for the $\nu=1/3$ state. Our study indicates that the $\nu=5/2$ and $7/3$ edge states are reconstructed for all experimental systems investigated so far and thus must be taken into account when analyzing experimental results. We also consider an effective field theory to gain insight into how edge reconstruction might influence various observable quantities.
\end{abstract}


\maketitle

\section{Introduction \label{sec:intro}}

When two-dimensional electrons are placed in magnetic field, fractional quantum Hall (FQH) states are observed~\cite{Tsui82}. These states are labeled by the filling factor $\nu$ defined as the number of electrons divided by the number of available single-particle states in each Landau level. The majority of FQH states occur at filling factors with odd denominators but the $\nu=5/2$ FQH state \cite{Willett87} is an exception whose nature is still not fully settled. The leading candidates are a chiral $p$-wave pairing state~\cite{Moore1991362,PhysRevLett.66.3205} of composite fermions~\cite{Jain89} and its particle-hole conjugate~\cite{Lee07, Levin07}. The Abrikosov vortex of a two-dimensional chiral $p$-wave superconductor supports a Majorana zero mode obeying non-Abelian braiding statistics~\cite{Read:2000zr, Ivanov01}, which is also predicted to be present in the 5/2 state~\cite{Moore1991362,Read:2000zr}. This article reports our study of the edge physics of the $\nu=5/2$ state. Our study has primarily been motivated by the proposals that interference of quasiparticles moving along the edge can reveal their non-Abelian nature~\cite{PhysRevLett.94.166802,PhysRevLett.96.016802,PhysRevLett.96.016803,PhysRevB.55.2331,Fradkin1998704}. In a more general context, FQH edge states have attracted attention because of the possibility of using them as a probe into the bulk topological order of FQH states~\cite{Wen:advances1995}. For example, the exponent characterizing the tunneling conductance into FQH edge has been predicted to depend only on the bulk topological properties~\cite{Wen:advances1995}. The experimental measurements for the FQH states at $\nu=n/(2n{\pm}1)$~\cite{Chang03} as well as the $5/2$ state~\cite{Marcus:2007np} have not yet yielded a quantized edge tunneling exponent, and several theoretical works have sought to shed light on the origin of the discrepancy~\cite{Conti96,Shytov98,Lopez99,Mandal01b,Mandal02b,Zulicke03,Wan05,Jolad07,Jolad09}. In particular, it has been found that the edge of a FQH state can undergo a reconstruction, and when that happens, it loses some of its universal features~\cite{Wan02, Wan03} (edge reconstruction for $\nu=1$ integer quantum Hall effect was considered in Ref.~\onlinecite{Chamon94}). This motivates us to seek a better and more detailed understanding of the edge physics for various FQH states, in particular of the $5/2$ FQH state, in realistic geometries.

In this work, we study the possibility of edge reconstruction at $\nu=5/2$ by modeling the confinement potential in a quasi-realistic manner as a uniform positively charged disk at a setback distance $d$ from the two-dimensional electron system. It is possible to carry out exact diagonalization for small systems~\cite{Wan06,Wan08} and such results are presented below for some cases. However, one may worry if these small system studies are able to capture the thermodynamic behavior of the true edge states. To access the edge excitations in larger systems, we use the trial wave functions for the ground state and edge excitations given by the Pfaffian model of the $5/2$ state~\cite{Moore1991362,PhysRevLett.66.3205,Wen:1993,PhysRevB.59.8084}. We study the edge excitations by diagonalizing the Coulomb interaction within the subspace spanned by the Pfaffian basis~\cite{Lee14}.  Based on comparisons with exact results obtained in small systems, we find that this model qualitatively captures the behavior of the actual system, although it slightly underestimates the critical separation $d_c$ beyond which edge reconstruction takes place. Our principal finding is that edge reconstruction occurs quite generically. For a model that completely disregards the lowest filled Landau level, edge reconstruction occurs for $d {\gtrsim} 0.5\ell_{B}$, where $\ell_B=\sqrt{{\hbar}c/eB}$ is the magnetic length. If the lowest filled Landau level is included, edge reconstruction occurs for $d{\gtrsim}0.1\ell_{B}$. This puts constraints on the experimental geometries where the physics of unreconstructed edge at $5/2$ may be investigated, and suggests that all current experimental realizations of the $5/2$ state are likely to have reconstructed edges.

We also ask what is the nature of instability at the edge. According to the Pfaffian model of the $5/2$ state, the edge excitations are built out of a bosonic mode and a fermionic mode. We deduce the dispersions for these modes from the many body energy spectra, and also show that certain trial wave functions provide reasonably good descriptions for them. The velocities of the bosonic and the fermionic modes are thus calculated from the dispersions. We find that it is the fermionic mode that goes soft first, rather than the bosonic mode as implicitly assumed in previous studies. We stress that this conclusion is based on a model that restricts to the Pfaffian basis of excitations; a similar analysis of the spectrum in the full basis has not been possible.

To investigate how edge reconstruction affects various experiments that involve the detailed behavior of the edge, we study the scaling behavior of electrons and quasiparticles at the reconstructed edge using a K matrix formalism. We find that the instability driven by the charged bosonic mode and that driven by the Majorana fermion mode lead to different fixed point behaviors. 

It is believed that the Pfaffian state and its particle-hole conjugate, namely the ``anti-Pfaffian" state, which is a Pfaffian of holes in the background of one filled Landau level (LL), exhibit topologically distinct edge structures~\cite{Lee07, Levin07}. A two-body interaction in the absence of LL mixing does not distinguish between the Pfaffian and the anti-Pfaffian insofar as their bulk properties are concerned, as would be the case in the torus geometry. However, on a finite disk with a boundary, the one-body potential breaks particle hole symmetry, and thus differentiates between the Pfaffian and the anti-Pfaffian, as explained in Sec. VII of Wan {\em et al.}\cite{Wan08}. We have considered below the edge reconstruction for the Pfaffian state. We stress that we have not included in our Hamiltonian a three-body interaction that may arise from LL mixing; for sufficiently small LL mixing this interaction should not affect the edge physics.

The $7/3$ FQH state has also been investigated in a number of recent experiments~\cite{Baer14,73exp1,73exp2,73exp3,73exp4,73exp5,73exp6,73exp7}.  We also study the edge of the $\nu=7/3$ state using exact diagonalization in the full Hilbert space (without LL mixing) as well as in the subspace given by the composite fermion (CF) theory. The trial wave functions given by the CF theory are not very accurate quantitatively for the $7/3$ ground state and excitations (see, {\em e.g.}, Ref.~\onlinecite{Balram13}), but are the best available model that can be dealt with in a simple manner. We find that, within this model, the edge of the $7/3$ state is also reconstructed for $d{\gtrsim}0.5\ell_B$. As the samples used in current experiments have larger setback distances, our results indicate that it is important to take edge reconstruction into account in the analysis of these experiments.

We stress the limitations of our calculation. We uncritically assume the Pfaffian model of excitations, which provides a restricted basis of edge excitations  -- our conclusions for the edge reconstruction are fully based on this assumption. Limited studies of edge reconstruction in the full basis (which can be performed in relatively small systems) suggest that this model is reasonable for obtaining the parameters where edge reconstruction occurs, but we are not able to provide a similar justification for the nature of edge reconstruction. Our calculation does not address, and thus does not rule out, the possibility that the experimental 5/2 state might be described by a model other than the Pfaffian model, such as the 331 or 113 state of Halperin \cite{Halperin83}, which has been supported by certain experiments \cite{Lin12,Baer14} and has been considered theoretically (see Ref. \onlinecite{Yang14b} and references therein). Similarly, for 7/3 FQHE, our conclusions are based on a model that restricts to the CF basis of edge excitations analogous to those at $\nu=1$. It is known that the excitations of 7/3 are, at least quantitatively, different from those at 1/3, and the theory that works very well for 1/3 excitations does poorly for 7/3 excitations \cite{Balram13b}; limited exact diagonalization studies indicate, however, that the model is not unreasonable for the question of edge reconstruction. Nonetheless, while our conclusions are quite reasonable for the stated models, the applicability of the models to the actual experimental system has not been fully confirmed. Finally, in Sec.~\ref{sec:exp} we  use an effective theory for edge states, and the validity of our results rests on the validity of the effective theory. 

The plan of the paper is as follows. In Sec.~\ref{sec:method}, we introduce the methods that are employed to calculate the energy spectra. We then describe our results for the $\nu=5/2$ FQH state in Sec.~\ref{sec:5/2} and discuss their implications for experiments in Sec.~\ref{sec:exp} in an effective field theory. The results for the $\nu=7/3$ FQH state are given in Sec.~\ref{sec:7/3} and finally, we conclude with some discussions in Sec.~\ref{sec:d&c}.

\section{Methods of Calculation \label{sec:method}}

Our numerical calculations will be performed using the disk geometry, with the single-particle wave functions given by
\begin{equation}
\begin{aligned}
\phi^n_m(x,y) &= \frac{1}{\sqrt{2\pi 2^{m+2n} n! (m+n)!}} \\
&\times \text{e}^{z\bar{z}/4} \left( -2\frac{\partial}{\partial z} \right)^{n} \left( -2\frac{\partial}{\partial \bar{z}} \right)^{m+n}  \text{e}^{-z\bar{z}/2}
\end{aligned}
\end{equation}
where $z=(x-iy)/\ell_B$ is the dimensionless complex coordinate, $n$ is the LL index, and $m$ labels the single particle orbitals within a LL. The lowest Landau level wave functions are particularly simple
\begin{eqnarray}
\phi^0_m(z) = \frac{z^m}{\sqrt{2\pi2^mm!}} \exp\left( -\frac{|z|^2}{4} \right)
\end{eqnarray}
This wave function is localized along a ring of radius $\sqrt{2m}$. The ubiquitous Gaussian factor will be omitted in the following discussions. 

\subsection{Microscopic model for exact diagonalization}

We consider a system consisting of two-dimensional electrons and a neutralizing background charge that is uniformly distributed on a disk with radius $R_{N}$ placed at a setback distance $d$ from the electron plane. For $\nu=5/2=2+1/2$, we treat the spin-up and spin-down electrons in the completely filled lowest Landau level (LLL) as inert and only use the Fock states of the half filled second Landau level (2LL). A confinement potential is provided by the background charge. To maintain overall charge neutrality, the total background charge is equal to that of the half-filled 2LL, which gives $R_{N}=\sqrt{4N}$ in unit of the magnetic length $\ell_B=\sqrt{{\hbar}c/eB}$. The Hamiltonian of this system is
\begin{equation}
\begin{aligned}
H =& E_K+V_{\rm ee}+V_{\rm eb}+V_{\rm bb} \\
=& \sum_i \frac{1}{2m_b} \left( \mathbf{P}_i + \frac{e}{c} \mathbf{A}_i \right)^2 + \sum_{i<j} \frac{e^2}{\epsilon| \mathbf{r}_i - \mathbf{r}_j|} \\
\quad -& \rho_0 \sum_i \int_{\Omega_N} d^2 {\bf r} \frac{e^2}{\epsilon\sqrt{|\mathbf{r}_i-\mathbf{r}|^2+d^2}} \\
\quad +& \rho_0^2 \int_{\Omega_N} d^2{\bf r} \int_{\Omega_N} d^2{\bf r^{\prime}} \frac{e^2}{\epsilon|\mathbf{r}-\mathbf{r}'|},
\label{eq1}
\end{aligned}
\end{equation}
where the terms on the right hand side represent the kinetic energy, electron-electron interaction, electron-background interaction, and background-background interaction, respectively. For electrons confined to the 2LL, the kinetic energy term is a constant and we only need to consider the interaction terms.

\subsection{Model wave functions}

In this paper, we use the so-called smooth edge model, which is believed to be relevant for the point-contact geometry used in experiments. In this model, all the possible many-body edge states for a given total angular momentum $M=M_{0}+{\Delta}M=\sum_{i=1}^{N}m_{i}$ ($M_{0}$ is the angular momentum of the ground state) are included with no constraints on the single-particle angular momentum $m_{i}$. The dimension of the Hilbert space grows exponentially with $N$. To access larger $N$, we will use model wave functions to generate a truncated subspace in the full Hilbert space, and diagonalize the Hamiltonian in this truncated subspace to obtain the energy spectra.  

\subsubsection{Composite fermion theory}

The FQH states in the lowest LL are described in terms of composite fermions~\cite{Jain89}, bound states of electrons and $2p$ vortices. As a first order approximation, composite fermions do not interact with each other and move in an effective magnetic field, forming Landau-like levels called $\Lambda$ levels. The CF filling factor $\nu^{*}$ corresponds to the electron filling factor $\nu=\nu^{*}/(2p\nu^{*}{\pm}1)$.  When $\nu^{*}=n$ is an integer, the composite fermions form a gapped integer quantum Hall state, which corresponds to an incompressible FQH state of the electrons at $\nu=n/(2pn{\pm}1)$. At the mathematical level, attachment of $2p$ vortices is accomplished by multiplication by the Jastrow factor $\prod_{i<j}(z_{i}-z_{j})^{2p}$. The CF wave functions are given by
\begin{equation}
\Psi^{\rm CF}_{\frac{\nu^{*}}{2p\nu^{*}+1}} = P_{\rm LLL} \Phi_{\nu^{*}} \prod_{i<j} \left( z_i - z_j \right)^{2p}
\end{equation} 
and 
\begin{equation}
\Psi^{\rm CF}_{\frac{\nu^{*}}{2p\nu^{*}-1}} = P_{\rm LLL} \left[ \Phi_{\nu^{*}} \right]^* \prod_{i<j} \left( z_i - z_j \right)^{2p}
\end{equation} 
where $\Phi_{\nu^{*}}$ is the Slater determinant wave function of non-interacting particles at filling factor $\nu^*$, $\left[ \cdots \right]^*$ denotes complex conjugate, and $P_{\rm LLL}$ is the LLL projection operator. The low-energy properties of the interacting electrons are accurately reproduced by the non-interacting composite fermions. The number of CF basis states at a particular angular momentum $M_{0}+{\Delta}M$ is much smaller than the dimension of the full Hilbert space, which allows us to study larger systems. It has been shown that the CF theory describes the edge excitations of the LLL FQH states very accurately~\cite{PhysRevB.51.9895,Mandal01b,Mandal02b}. In this paper, we will construct edge excitations for the $1/3$ state by mapping it to composite fermions at $\nu^*=1$, and explore the edge excitations of the $7/3$ state using an effective interaction mimicking the second LL Coulomb interaction.  

\subsubsection{Pfaffian state and Jack polynomials}

The trial wave function for the $5/2$ state that we consider in this paper is the Pfaffian state~\cite{Moore1991362}:
\begin{equation}
\begin{aligned}
\Psi ^{\rm Pf}_{1/2}\left( \{z\} \right) = {\rm Pf} \left( \frac{1}{z_i-z_j} \right) \prod _{i<j} (z_i - z_j )^2,
\label{eq2}
\end{aligned}
\end{equation}
which represents a chiral $p$-wave paired state of composite fermions. This wave function is written in the LLL at filling factor $1/2$ and it is the highest density zero energy eigenstate of a model 3-body interaction Hamiltonian~\cite{PhysRevLett.66.3205,Wan08}:
\begin{eqnarray}
H_3 &=& \sum_{i<j<k} S_{ijk} \Big[ \nabla^2_i \nabla^2_j \left( \nabla^2_i + \nabla^2_j \right) \nonumber \\
      &\times& \delta(\mathbf{r}_i - \mathbf{r}_j) \delta(\mathbf{r}_i - \mathbf{r}_k) \Big]
\end{eqnarray}
where $S_{ijk}$ is a symmetrization operator. Numerical studies have shown support for this interpretation of the $5/2$ state: the overlap of the trial wave function with the exact Coulomb ground state at $\nu=5/2$ is about $80\%$ for $18$ electrons on sphere~\cite{PhysRevLett.104.076803}, and the Pfaffian wave function has a lower energy than the spin-polarized or the spin-unpolarized composite fermion Fermi sea state in the 2LL~\cite{PhysRevB.58.R10167}. The edge excitations can be constructed following Ref.~ \onlinecite{PhysRevB.59.8084}, which are also zero energy eigenstates of the three-body Hamiltonian $H_3$. 

It is possible to diagonalize the three-body Hamiltonian $H_3$ to obtain the Pfaffian basis states. A simpler way is to use the Jack polynomial formalism~\cite{Bernevig08, Bernevig09,Thomale11,Lee14} which we briefly explain here. This approach gives explicit decomposition of certain model FQH wave functions in the Slater determinant basis. To begin with, we label the single-particle orbitals in LLL by their angular momentum eigenvalues. A non-interacting $N$-particle basis state, which has a fixed total angular momentum, can be represented by a partition $\lambda=[\lambda_{1}, \lambda_{2}, ..., \lambda_{N} ]$ with the angular momentum $\lambda_{i}$ of each particle listed in descending order, or an occupation number configuration $n(\lambda)=\{ n_{m}(\lambda), m= 0,1,2,... \}$ showing the number of particles $n_{m}$ in the single-particle state $m$. An useful operation on the many-body basis called ``squeezing" is defined as follows: when two orbitals $m_{1}$ and $m_{2}$ ($m_{1}<m_{2}-1$) are occupied, the elementary squeezing operation moves one particle in each orbital to the orbitals $m_{1}+1$ and $m_{2}-1$ (the Pauli principle should be satisfied when dealing with fermions). In terms of occupation numbers, we have $n_{m_{1}} \to n_{m_{1}}-1$, $n_{m_{2}} \to n_{m_{2}}-1$, $n_{m_{1}+1} \to n_{m_{1}+1}+1$, and $n_{m_{2}-1} \to n_{m_{2}-1}+1$. If a partition $\mu$ can be generated by squeezing another partition $\lambda$, we say that $\lambda$ dominates $\mu$ as denoted by $\lambda>\mu$. The decomposition of a fermionic Jack polynomial in terms of Slater determinants contains only the partitions dominated by a certain ``root partition" as follows: 
\begin{equation}
S_{\lambda}^{\alpha} (z_{1}, ..., z_{N}) = \sum_{\mu \leq \lambda} b_{\lambda \mu} \text{sl}_{\mu}.
\end{equation}
Here, $\lambda$ denotes the root partition and the coefficients $b_{\lambda\mu}$ are determined recursively as
\begin{equation}
b_{\lambda \mu}= \frac{2(\frac{1}{\alpha}-1)}{\rho_{\lambda}^{F} (\alpha) - \rho_{\mu}^{F} (\alpha)  } \sum_{\theta ; \mu <\theta \leq \lambda} (\mu_{i} - \mu_{j}) b_{\lambda \theta} (-1)^{N_{\rm SW}},
\label{eq:jack}
\end{equation}
where $\rho_{\lambda}^{F} (\alpha) = \sum_{i} \lambda_{i} [\lambda_{i} +2i(1-1/\alpha)]$ and the parameter $\alpha=-3$ for the Pfaffian state. The sum in equation (\ref{eq:jack}) extends over all partitions $\theta = [\mu_{1}, ..., \mu_{i} +s, ..., \mu_{j}-s, ..., \mu_{N}]$ that strictly dominate the partition $\mu=[\mu_{1}, ... , \mu_{N}]$ and are squeezed from the root partition $\lambda$. $N_{\rm SW}$ is the number of swappings that are needed to bring the partition $\theta$ back to ordered form. The root configuration $\lambda$ implements a ``generalized Pauli principle'', which for the Pfaffian state requires no more than $2$ particles in $4$ consecutive orbitals. This helps us to determine the root configurations for the ground state as well as the edge excitations. For example, in the occupation number picture, the Pfaffian ground state with $N=6$ particles has root configuration $[1100110011]$, and there is only one possible edge state at $\Delta M=1$ with root configuration $[11001100101]$. The expansion coefficients of the ground state and edge excitations can be calculated using Eq.~(\ref{eq:jack}) when their root configurations are known. Since each Slater determinant corresponds to a second quantized many-body basis state, the Jack polynomial formalism gives both real space expressions and second quantized state vectors in Fock space.

\subsubsection{Bosonic and fermionic edge mode wave functions}

It has been postulated that the edge excitations of the Laughlin states are described by a chiral bosonic mode~\cite{Wen:advances1995}, while the edge excitations of the Pfaffian state contain a chiral bosonic mode and a chiral Majorana fermionic mode~\cite{Wen:1993}. We will try to identify the edge modes in the numerically obtained energy spectra and also construct trial wave functions for the edge excitations to gain further insight into their nature. 

Oaknin, Martin-Moreno, Palacios, and Tejedor (OMPT) \cite{PhysRevLett.74.5120} introduced the operators
\begin{equation}
{\widehat S}_k^{\dagger}=\sum_{n=0}^{\infty} \sqrt{\frac{n!}{(n+k)!}} c_{n+k}^\dagger c_n 
\end{equation}
to create edge excitations. Applying ${\widehat S}_k^{ \dag}$ on a LLL wave function increases its total angular momentum by $k$, which corresponds to the presence of edge excitations. It can be shown that ${\widehat S}_k^{\dagger} |\psi_1\rangle$ ($\psi_{1}=\prod_{i<j}{(z_i-z_j)}$ is the $\nu=1$ integer quantum Hall state) is an eigenstate of the center of mass angular momentum operator. This makes the OMPT operator a better choice compared to a simple density operator. 

To construct trial wave functions for the single-boson edge mode of the $\nu=1/3$ or the $7/3$ state at angular momentum $\Delta M$ (measured with respect to the ground state), we multiply the ${\widehat S}_{\Delta M}^{\dagger}|\psi_1\rangle$ with the Jastrow factor $\prod_{i<j} (z_i-z_j)^{2}$ and rewrite this state as
\begin{equation}
\Psi_{1/3}^{\Delta M} (\{z\}) = \frac{{\widehat S}_{\Delta M}^{\dagger} |\psi_1\rangle}{|\psi_1\rangle} \Psi^{\rm GS}_{1/3}(\{z\}), 
\label{eq:OMPTCF}
\end{equation}
where $\Psi^{\rm GS}_{1/3}(\{z\})=\prod_{i<j} (z_i-z_j)^{3}$ is the Laughlin $1/3$ ground state. It has been demonstrated that all the edge excitations of the $1/3$ state can be created by populating a single bosonic mode which corresponds to the lowest branch in the energy spectra~\cite{Jolad:2010}. As we will show below in Sec. V, the results at $\nu=7/3$ are similar to those at $1/3$.  

We can generalize the OMPT method to create the bosonic edge excitations for the Pfaffian state as
\begin{equation}
\Psi_{\rm Pf}^{\Delta M} (\{z\}) = \frac{{\widehat S}_{\Delta M}^{\dagger} |\psi_1\rangle} {|\psi_1\rangle} \Psi_{\rm Pf}(\{z\}).
\label{eq:PfOMPT}
\end{equation}
In Sec. III, we will compare the trial wave functions with the single-boson edge mode extracted from the full energy spectra. 

For the Majorana fermionic edge mode of the Pfaffian state, Milovanovi\'c and Read~\cite{PhysRevB.53.13559} proposed the wave function
\begin{eqnarray}
&& \Psi_{n_1,...,n_F}(z_1,...,z_N) = \frac{1}{2^{(N-F)/2}(N-F)/2!} \nonumber\\
&\times& \sum_{\sigma \in S_N}{ \mathrm{sgn} \sigma \frac{\prod_{k=1}^F{z_\sigma(k)^{n_k}}}{( z_{\sigma(F+1)} - z_{\sigma(F+2)} )...( z_{\sigma(N-1)} - z_{\sigma(N)} )}} \nonumber \\
&\times& \prod_{i<j} (z_i - z_j )^2 
\label{eq:MR}
\end{eqnarray}
at 
\begin{equation}
\Delta M= \sum_{k=1}^F\left(n_k+ \frac{1}{2}\right).
\end{equation}
This wave function is interpreted as having $F$ fermions created in angular momentum orbitals $\Delta M=n+1/2$ with $n=0,1, \cdots$. The evaluation of this wave function is difficult because of the antisymmetrization operator; the maximum number of electrons that we are able to reach is $10$. We also compare the trial wave functions with the pure Majorana fermionic edge modes that we can identify in the full energy spectra in Sec. III.

\subsection{Real space wave function and effective interaction}

For each angular momentum value, the model wave functions that we have introduced in the previous subsection define a truncated subspace and the Coulomb Hamiltonian can be diagonalized within this subspace. To be specific, suppose that we know the real space wave functions \{$\Psi_{\alpha}^{\Delta M}(\{z\})$\} at relative angular momentum ${\Delta}M$, with $\alpha$ labeling the different states. We can then evaluate the Coulomb matrix elements $V^{{\Delta}M}_{\alpha\beta}=\langle\Psi_{\alpha}^{\Delta M}|V|\Psi_{\beta}^{\Delta M}\rangle$ (a multi-dimensional integral) using Metropolis Monte Carlo algorithm \cite{Metropolis:1953}. Because the basis wave functions are in general not orthogonal, the Gram-Schmidt method should be applied to find the Coulomb matrix in the orthonormal basis (see, {\em e.g.} Ref.~\onlinecite{Jain07}), which can be diagonalized to find the energy eigenvalues. The dimension of this truncated subspace is significantly smaller than that of the full Hilbert space. Thus, if the trial wave functions can be evaluated efficiently in real space, this method can be used to explore systems larger than those accessible to the exact diagonalization approach. Of course, the accuracy of the results depends on the accuracy of the basis. 

We note that the trial wave functions presented above are written in the LLL, but we are interested in the $5/2$ and $7/3$ FQH states which occur in the 2LL. One may attempt to convert the LLL wave functions to their 2LL counterparts, which is not only difficult in practice and but also undesirable because of the complexity of the 2LL wave functions. One can alternatively use the LLL wave functions with an effective interaction to mimic the physics in the 2LL. For the electron-electron interaction $V_{\rm ee}$, an effective interaction of the form 
\begin{equation}
V_{\text{eff}}(r)=\frac{1}{r}+\frac{1}{\sqrt{r^6+1}}+\frac{9}{4\sqrt{r^{10}+10}}+\sum_{k=0}^{K-1}C_{k}r^{2k} e^{-r^2},
\end{equation}
has been proposed~\cite{ChuntaiShi:2008}, which is determined by demanding that it has the same Haldane pseudopotentials in the LLL as the Coulomb interaction does in the 2LL. The electron-background interaction $V_{\rm eb}$ should also be replaced by an effective interaction as explained in Appendix~\ref{appx:eb}. If the setback distance $d{\neq}0$, the effective interaction is given by 
\begin{widetext}
\begin{equation}
V_{\rm eb}^{(\text{eff})}({\bf r})= -\frac{e^2 \rho_0}{\epsilon} \int_{\Omega_N}{d^2 {\bf r^\prime}  \left( \frac{1}{ R} + D_2 \frac{1}{ R^2}+D_3 \frac{1}{ R^3} +D_4 \frac{1}{R^4} +D_5 \frac{1}{R^5} \right)},
\label{eq:Vebeffform}
\end{equation}
\end{widetext}
where $R ({\bf r, r^\prime},d)=\sqrt{| {\bf r}-{\bf r^\prime} |^2+d^2}$ and the values for $D_k$ are given in Table~\ref{tab:D}. The integration in Eq.~(\ref{eq:Vebeffform}) has a singularity at $d=0.0$, so we need to use a more complicated form which in this case is given by
\begin{widetext}
\begin{equation}
V_{\rm eb}^{(\text{eff})}(r)=-\frac{e^2 \rho_0}{\epsilon} \int_{\Omega_N}{d^2 {\bf r^\prime}  \left( \frac{1}{R}+\frac{1}{2\sqrt{R^6+1}}+\sum{C_{i} R^i e^{-\beta_{i} R^{2}}} \right)}, \label{deq0}
\end{equation}
\end{widetext}
where $R=R ({\bf r, r^\prime})=| {\bf r}-{\bf r^\prime} |$. We find that a convenient form is obtained with the choice $\beta_i=6$ and $C_0=-594.631, C_1=4137.098, C_2=-7882.778,C_3=4457.804$, $C_4=C_5= \cdots =0$.

\begin{table}
\begin{ruledtabular}
\begin{tabular}{c|cccc}
$\quad  d \quad $   & $D_2=D_4$ & $D_3$ &  $D_5$ \\ \hline
0.5\quad 	&0.0 			&0.5			&-0.375\\						
1.0\quad 	&0.0 			&0.5			&-1.5\\						
1.5\quad 	&0.0 			&0.5			&-3.375\\		
2.0\quad 	&0.0 			&0.5			&-6.0				
\end{tabular}
\end{ruledtabular}
\caption{$D_k$ Values for $d=0.5-2.0$}
\label{tab:D}
\end{table}

For the $\nu=7/3$ state, the trial wave functions can be evaluated easily using Monte Carlo method in real space, so we can study relatively large systems using the effective interaction approach. The energies of the OMPT trial wave functions for the single-boson edge modes can also be found in a similar way. In contrast, for the Pfaffian state, real space wave functions given by the Jack polynomial formalism are linear superpositions of a large number of Slater determinants and much more computational time would be needed if one uses the Monte Carlo method. We therefore use a different method for the $5/2$ state as explained in the next subsection. 

\subsection{Second quantized approach}

Another method to obtain the energy spectrum is using the second quantized form of the Hamiltonian. The single-particle angular momentum eigenstates within a LL are labeled by integers, and the Fock states in the many-body Hilbert space can be written as $|m_{1},m_{2},...,m_{N}\rangle$. The Hamiltonian can be expressed in second quantized form as in Eq.~(\ref{eq:2ndqH}) and the matrix elements in the Fock state basis can be evaluated. Diagonalizing this Hamiltonian matrix gives the full energy spectrum, but this becomes impractical for large systems due to the exponential growth of the Hilbert space dimension. To get access to larger systems, we restrict ourselves to the truncated subspace within the Pfaffian model of the $5/2$ state. At each angular momentum, a few basis states describing the edge excitations of the $5/2$ state are generated using the Jack polynomial formalism. We construct the Coulomb interaction matrix in this truncated subspace and compute its eigenvalues. Using this method, we are able to study the edge excitations at $\nu=5/2$ for systems with up to $N=16$ electrons, while the largest system size that has been studied~\cite{Wan06} using exact diagonalization is $N=12$.

\section{Edge spectra and reconstruction at $\nu=5/2$ \label{sec:5/2}}

In this section, we present the energy spectra of different systems, which reveal the existence of edge reconstruction in certain parameter regime. We employ two different approximations, called Model I and Model II, in our calculations. In Model I, we neglect the completely filled LLL and only consider the half-filled second LL in the presence of a uniform neutralizing background. In Model II, the electrons in the LLL are also taken into account in a static manner, treated as a part of the neutralizing background; we thus have two neutralizing backgrounds in Model II: one made of donors at a setback distance $d$ and the other consisting of the LLL electrons that are spatially coincident with the electrons in the second LL. The Model II is more realistic. It should be noted, however, that we have not considered the possibility of edge reconstruction in the lowest LL; this may be justified from the expectation that the integer quantum Hall states are more robust, and thus less prone to edge reconstruction, than the FQH states.

\subsection{Model I for the $\nu=5/2$ edge}
	
\subsubsection{Small system study}

\begin{figure}
\hspace{-3mm}\resizebox{0.4\textwidth}{!}{\includegraphics{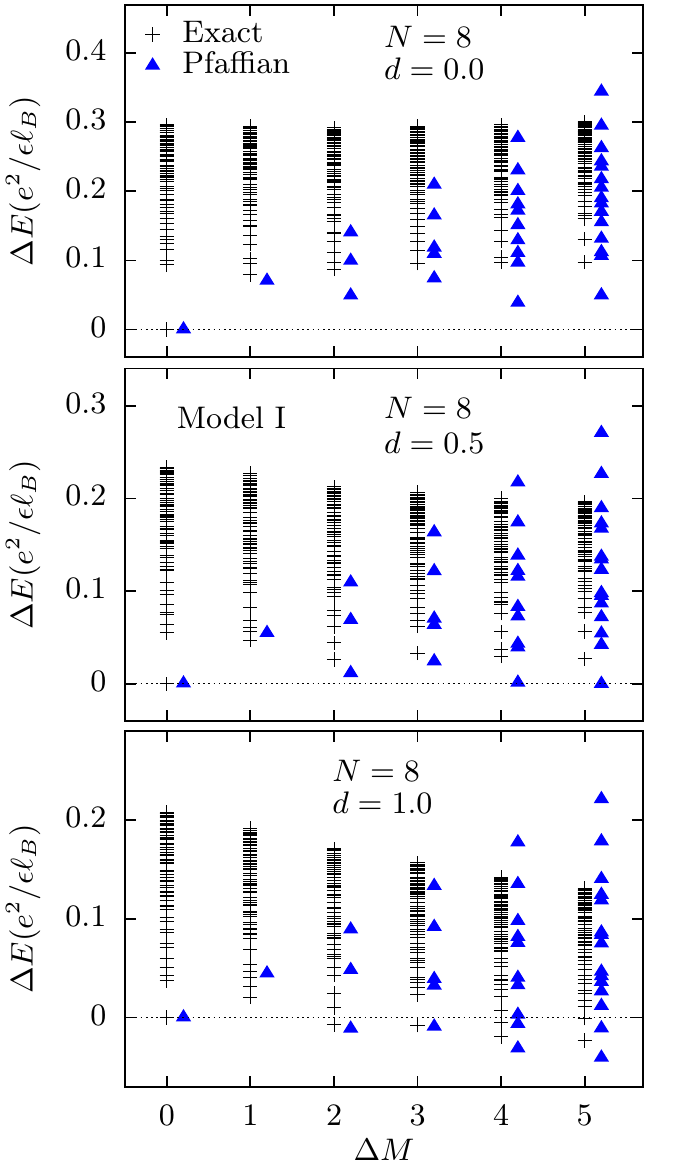}} \\
\vspace{-4mm}
\caption{(Color online) Comparison of the edge spectra obtained by diagonalization of the Model I Coulomb interaction within the full basis (black pluses) and the Pfaffian basis (blue triangles) for $N=8$ particles. The energies are measured relative to the ground state energy in both spectra. Similar overall trends are seen for the lowest spectral branch.}
\label{fig:exact}
\end{figure}

To test the validity of the Pfaffian model for the edge excitations of the $5/2$ state, we first perform exact diagonalization in the full Hilbert space for small systems and compare the results with those obtained within the Pfaffian basis. Fig.~\ref{fig:exact} shows the energy spectra of the $N=8$ system at different setback distances $d$. As one can see from the figure, the Pfaffian model captures the basic features of the low-energy part of the exact results. At $d=0.0$, no edge reconstruction is found, while both the full energy spectrum and the Pfaffian model exhibit edge reconstruction at $d=1.0$. A careful inspection of the results at various different values of $d$ shows that the Pfaffian model slightly overestimates the tendency toward edge reconstruction. With this finding in mind, we will use the Pfaffian model to study the edge excitations of the Pfaffian state in larger systems in the next subsection.

\begin{figure*}
\resizebox{0.85\textwidth}{!}{\includegraphics{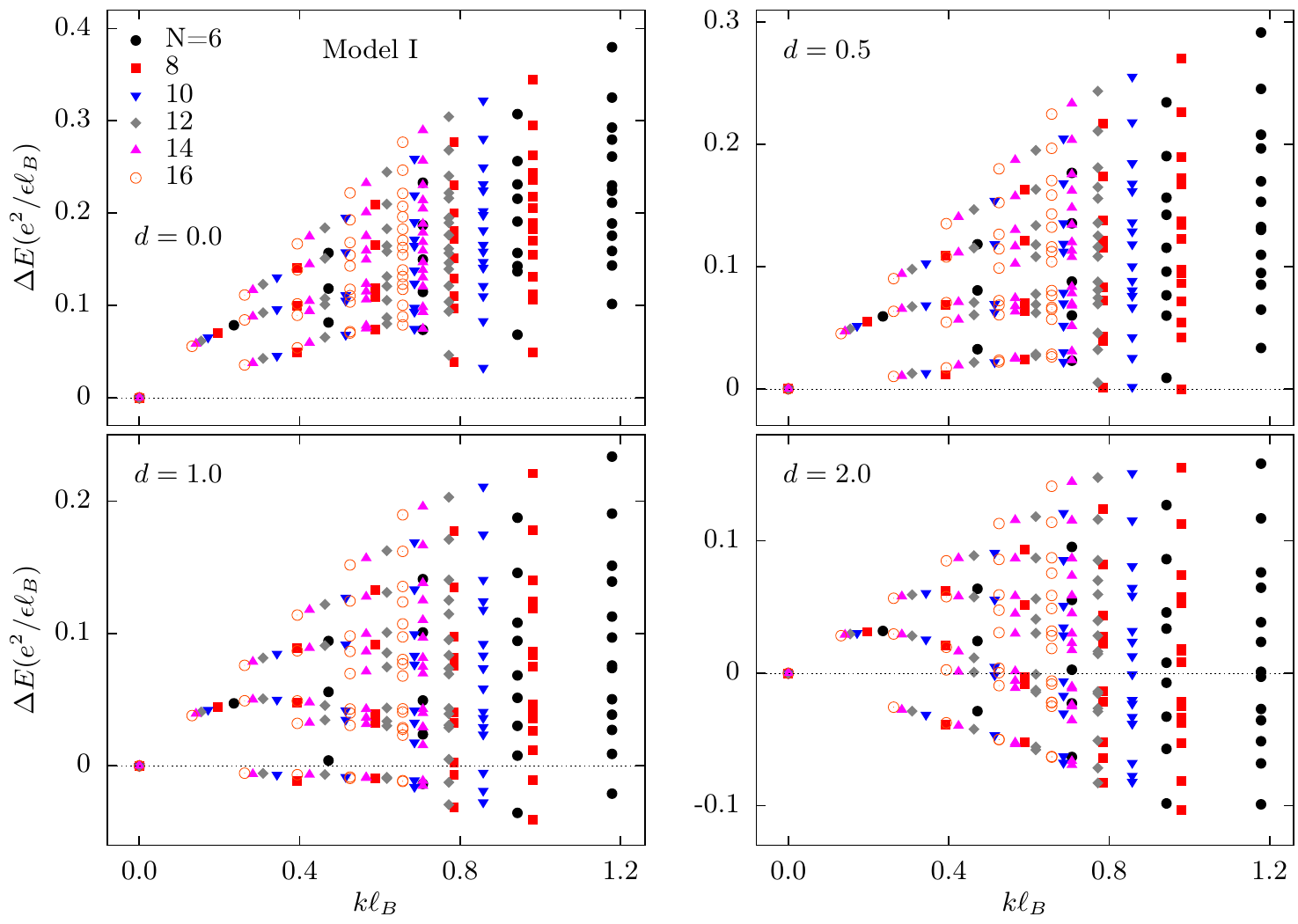}}\\
\vspace{-4mm}
\caption{(Color online) Edge spectra of $5/2$ state as a function of the physical momentum $k$ for $N=6-16$ particles. The energies are obtained by diagonalizing the full Hamiltonian of Model I within the Pfaffian edge basis. Data from $N=8-16$ collapse into a single curve for the lowest spectral branch. Edge reconstruction occurs when $d \geq 0.5\ell_{B}$.}
\label{fig:pfaf}
\end{figure*}

\subsubsection{Pfaffian edge spectra and edge reconstruction}

Within the truncated subspace given by the Pfaffian model, we are able to compute the energy spectra of systems with $N{\leq}16$ particles for $\Delta M=0-5$. The energy spectra corresponding to different system sizes are compared to test whether the thermodynamic limit has been reached. For this purpose, we need to use a scaling relation between the angular momentum $\Delta M$ and the physical momentum $\delta k$ (we choose $\hbar=1$ for simplicity), which can be obtained from the expression of the system size in terms of $\Delta M$ and $\delta k$. For a system with the Pfaffian ground state, the radius of wave function with relative angular momentum $\Delta M$ is $r=\sqrt{2M}\ell_B=\sqrt{2(2(N-1)-1+\Delta M)}\ell_B$, while the physical momentum is related to the size of the system via $k{\sim}r/\ell^2_B$. This leads to the following definition for the momentum of the edge excitation
\begin{equation}
\delta k = \frac{\Delta M} {\sqrt{4N-6}}\frac{1}{\ell_B}.
\label{Eq:Mkrelation}
\end{equation}
We plot the energy spectra versus the physical momentum $\delta k$ (simply denoted as $k$ in what follows) in Fig.~\ref{fig:pfaf} for systems with $N=6-16$, $d=0-2.0\ell_B$, and $\Delta M=0-5$.

It can be seen from Fig.~\ref{fig:pfaf} that the lowest energy branches of different systems collapse to a single curve for $N=8-16$ while the second lowest branches collapse on to a single curve for $N=10-16$, indicating that we have achieved proper scaling to the thermodynamic limit, and thus ascertaining the validity of studying the real system using the currently available system sizes. As the setback distance increase, the confinement potential gets weaker and edge reconstruction occurs when $d>d_{c}{\approx}0.5\ell_B$. This is in good semiquantitative agreement with the critical setback distance of $0.5-0.8 \ell_{B}$ found in Ref. \onlinecite{Zhang13}. 

\subsubsection{Dispersion of bosonic and fermionic modes}

\begin{figure*}
\hspace{-4mm}\resizebox{0.85\textwidth}{!}{\includegraphics{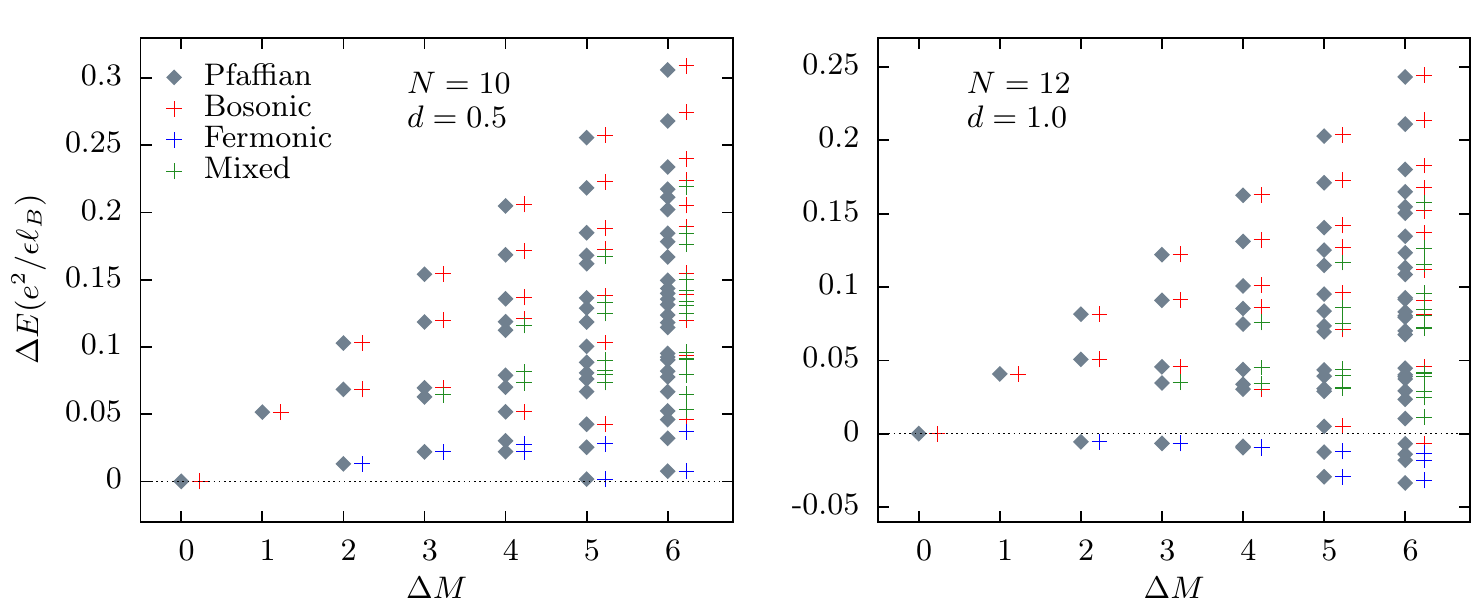}}\\
\vspace{-4mm}
\caption{(Color online) Bosonic and fermionic modes in the Pfaffian spectra. The grey diamonds indicate the Coulomb spectrum as obtained by exact diagonalization within the Pfaffian basis. We obtain the energy dispersion of the single boson and single fermion excitations using the exact spectra, from which the full spectrum can be built with the assumption that these modes are non-interacting; this spectrum is indicated by pluses. }
\label{fig:modes}
\end{figure*} 

It was proposed that the edge excitations of the Pfaffian state consist of a chiral bosonic mode plus a chiral Majorana fermionic mode~\cite{Wen:1993}. To analyze these two modes, we label each edge excitation by $n_b(l_b)$ and $n_f(l_f)$, which are the occupation numbers of the bosonic and fermionic modes at angular momenta $l_b$ and $l_f$ with energies $\epsilon_b$ and $\epsilon_f$, respectively. The quantity $n_b(l_b)$ can be any non-negative integers while $n_f(l_f)$ is $0$ or $1$. The angular momentum $l_b$ for the bosonic mode must be an integer, while $l_f$ for the Majorana fermionic mode must be a half odd integer due to antiperiodic boundary condition~\cite{Wen:1993}. If we assume that the edge modes are non-interacting, the angular momentum and energy of the state labeled by $n_b(l_b)$ and $n_f(l_f)$ are, measured with respect to the ground state,
\begin{equation}
\begin{aligned}
\Delta M &=\sum_{l_b}n_b(l_b)l_b+\sum_{l_f}n_f(l_f) l_f , \\
\Delta E &=\sum_{l_b}n_b(l_b)\epsilon_b(l_b)+\sum_{l_f}n_f(l_f)\epsilon_f(l_f),
\label{eq:AngEne}
\end{aligned}
\end{equation}
respectively. Thus, given the dispersions $\epsilon_b(l_b)$ and $\epsilon_f(l_f)$ for a single boson and a single fermion, we can construct the full spectrum containing many bosons and fermions. (Fermions appear only in even numbers due to their Majorana nature.) 

\begin{figure}
\resizebox{0.42\textwidth}{!}{\includegraphics{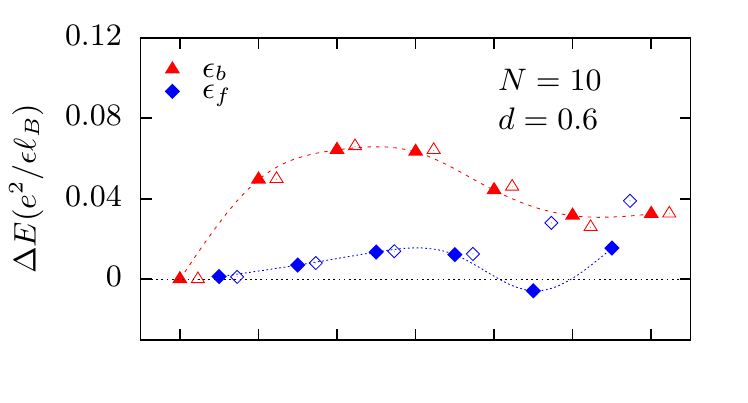}}\\
\vspace{-9.3 mm}
\resizebox{0.42\textwidth}{!}{\includegraphics{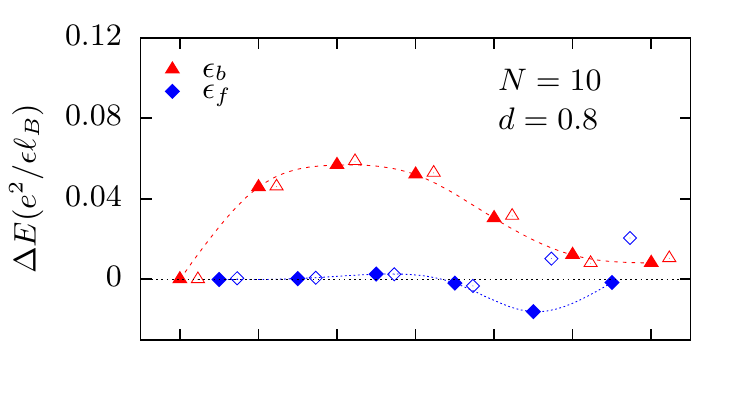}}\\
\vspace{-9.3 mm}
\resizebox{0.42\textwidth}{!}{\includegraphics{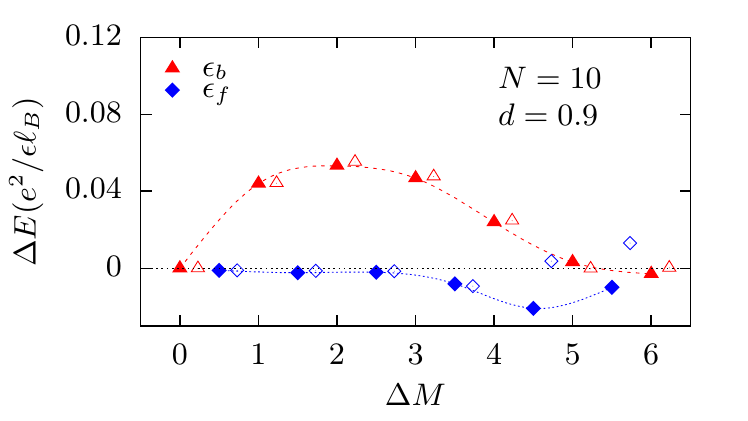}}\\
\vspace{-4.0 mm}
\caption{(Color online) Energy dispersion curves for bosonic (red triangles) and fermionic (blue diamonds) modes for $5/2$ system with $N=10$ particles at different values of $d$. The solid symbols are extracted from the full energy spectra in the Pfaffian subspace while the empty blue and red symbols are from the MR and OMPT model wave functions given in Eqs.~\ref{eq:MR} and \ref{eq:OMPTCF}, respectively. }
\label{fig:PfmodeE}
\end{figure}

We deduce the single particle dispersions $\epsilon_b(l_b)$ and $\epsilon_f(l_f)$ from the spectrum obtained by diagonalizing the Coulomb interaction in the Pfaffian basis, indicated by grey diamonds in Fig.~\ref{fig:modes}. The procedure is as follows.
For $\Delta M=1$, there is only one state which we identify as the bosonic state with energy $\epsilon_{b}(1)$. The highest-energy state at $\Delta M=2$ is evidently the two boson state $\Delta E = 2\epsilon_{b}(1)$. There are two additional states left at $\Delta M  = 2$ with energies $\epsilon_{b}(2)$ and $\epsilon_{f}(1/2) + \epsilon_{f}(3/2)$. We identify the lower energy state as the fermionic mode and the higher energy one as the bosonic mode. (As a result, the low lying states at higher $\Delta M$ will also be identified with pure fermionic modes.) This identification is justified from the following two observations. (i) We will see that we are able to give an excellent account of the full spectrum in terms of the spectrum predicted by Eq.~(\ref{eq:AngEne}); that would not be the case if we had assumed the bosonic mode to lie at lower energy. (ii) We also compute the spectra from the trial wave function for pure bosonic and fermionic modes, as we will show later; these are also consistent with the above identification. For $\Delta M = 3$, there are five states. We assign the lowest-energy state with the pure fermionic state with energy $\epsilon_{f}(1/2) + \epsilon_{f}(5/2)$. We can also easily identify the two states with highest energies as bosonic states $3 \epsilon_{b}(1)$ and $\epsilon_{b}(1)+\epsilon_{b}(2)$. From convolution of both bosonic and fermionic modes, we can find one more state with $\Delta E=\epsilon_{b}(1)+\epsilon_{f}(1/2)+\epsilon_{f}(3/2)$. The only edge state left is the single-boson state $\epsilon_{b}(3)$. Similar analysis for larger $\Delta M$ allows us to determine the dispersion relation of the bosonic mode uniquely in this way.  There is sometimes uncertainty in determining $\epsilon_{f}(l_{f})$ values. For example, as we go to $\Delta M=4$, there are two possible ways of creating a pair of fermionic excitations  with energies given by $\epsilon_{f}(1/2)+\epsilon_{f}(7/2)$ and $\epsilon_{f}(3/2)+\epsilon_{f}(5/2)$, respectively. We find, however, that one of those choices gives better agreement with the spectrum at larger $\Delta M$. In this manner, we are able to determine the dispersion relation $\epsilon_{f}(l_{f})$ which best reproduces the full original spectra containing multiple bosonic and fermionic excitations. The obtained dispersion relations for bosonic and fermionic modes are shown in Fig.~\ref{fig:PfmodeE} with solid shapes. We also reproduce the spectrum with non-interacting bosons and fermions according to Eq.~(\ref{eq:AngEne}) as shown with pluses in Fig.~\ref{fig:modes} for comparison with the original Pfaffian subspace energy spectrum.
The red pluses are pure bosonic modes; the lowest branches correspond to the single-boson excitations while the other red pluses to states containing multiple bosonic excitations. The blue pluses indicate edge states with pairs of pure fermionic excitations (pairs are needed to produce the physical integral angular momenta). The green plus show the mixed states containing both bosonic and fermionic excitations. The agreement between the pluses and the diamonds demonstrate that neglecting the interaction between the fundamental excitations (bosons or fermions) is a valid approximation, at least for small values of $\Delta M$ (and within a model that retains only the Pfaffian wave functions). This also gives us confidence in our assignment of the various modes in terms of fermionic, bosonic or mixed modes. The excellent agreement also demonstrates the quantitative reliability of the dispersions for the single boson and fermion modes within the assumed model.

One may ask if a similar assignment may be made using the full Coulomb spectra shown in Fig.~\ref{fig:exact}. Unfortunately, that is not possible due to the rather closely spaced nature of the eigenstates and possible mixing with other states. Our conclusions below are drawn from calculations within the Pfaffian basis. One may expect that the full Coulomb spectrum will also show this behavior for sufficiently large systems, but we are not able to confirm that. Wan {\em et al.} \cite{Wan08} have also noted that no gap separates edge and bulk states at this system size ($N=12$) for pure Coulomb interaction, and overlap calculation indicates that the edge modes mix with bulk excitations, thereby precluding a meaningful evaluation of the bosonic and fermionic dispersions.

We have also tested the OMPT wave function in Eq. (\ref{eq:PfOMPT}) and the Milovanovi\'c-Read (MR) wave function in Eq. (\ref{eq:MR}) for the single-boson and the pure fermionic pair edge excitations, respectively. Fig.~\ref{fig:PfmodeE} shows that the single-boson energies calculated directly from the OMPT functions (the empty red triangles) agree very well with the single boson dispersion obtained above from the Pfaffian subspace energy spectra (the solid red triangles). For the fermionic mode, the energies obtained using the MR wave functions are for pairs of fermionic excitations (or even multiple fermions when $\Delta M$ is large enough). We determine the energies $\epsilon_{f}(l_{f})$ of the single-fermionic excitations using similar analysis as we performed for the Pfaffian subspace spectrum. We then compare the two sets of $\epsilon_{f}(l_{f})$ values obtained from MR wave function and Pfaffian subspace spectrum in Fig.~\ref{fig:PfmodeE}, shown with empty and solid blue triangles respectively. The MR wave functions produce the same $\epsilon_{f}(l_{f})$'s as those extracted from the Pfaffian subspace energy spectra for small angular momenta, but tend to give higher energy when $\Delta M$ increases. The value of $\Delta M$ where the discrepancy becomes noticeable increases with $N$. For example, while the first four blue data points agree well in Fig.~\ref{fig:PfmodeE} for $N=10$, only the first three $\epsilon_{f}(l_{f})$ match well when $N=8$ (not shown here). The mismatch in small systems or for large $\Delta M$ is thus likely due to finite size effects. We have found that when we construct the full spectrum using the dispersions obtained from the OMPT and MR wave functions, the agreement with the actual spectrum is less satisfactory than that in Fig.~\ref{fig:modes}. Nonetheless, even the OMPT and MR dispersions indicate that the fermionic edge mode goes soft before the bosonic one.

In what follows, we will use the single boson and fermion dispersions obtained from the full Pfaffian spectrum, and the conclusions below are based on the spectrum produced by the Pfaffian basis.

The results obtained from the Pfaffian subspace energy spectra shed some new light into the edge reconstruction at $5/2$. Fig.~\ref{fig:PfmodeE} shows the dispersions of the single boson and the single fermion excitations for several values of $d$ for $N=10$ particles. 
Edge reconstruction occurs in all the three panels of Fig.~\ref{fig:PfmodeE}. Furthermore, the fermionic mode has the lowest energy, suggesting that it is the one that drives edge reconstruction. Indeed, in the full spectrum in Fig.~\ref{fig:modes} the lowest energy state is built from pure fermionic excitations. These results present a scenario of the edge modes and edge reconstruction that is somewhat different from the one suggested previously~\cite{Wan08} and addressed more thoroughly in Ref. \onlinecite{Zhang13}, which concluded, based on the orbital occupation number of the destabilizing state, that edge reconstruction occurs in the bosonic branch. Below we discuss how these two scenarios lead to different predictions for the edge exponents.
 
\begin{figure}
\resizebox{0.42\textwidth}{!}{\includegraphics{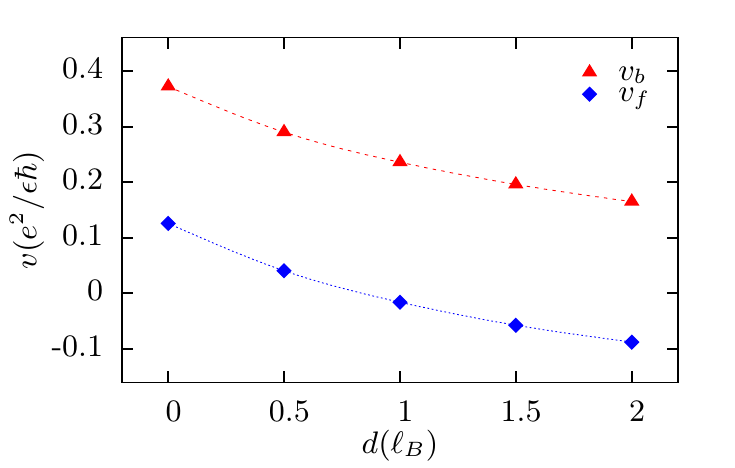}}\\
\vspace{-4.0mm}
\caption{(Color online) The velocities for the bosonic (red triangles) and fermionic (blue rhombi) modes as functions of $d$. Here, $N=10$ particles.}
\label{fig:velocity}
\end{figure}

 We have also calculated the velocity $v=d\epsilon/dk$ of each mode using Eq.~(\ref{Eq:Mkrelation}). We assume that the bosonic dispersion is linear for $\Delta M < 1$ and fit the $l_{f} \leq 5/2$ part of the fermionic dispersion using a straight line with zero intercept. The velocities of the bosonic and fermionic modes for each $d$ are shown in Fig.~\ref{fig:velocity} for $N=10$. We expect that similar values would be found in other systems because data collapses for both the bosonic and fermionic modes have been achieved in Fig.~\ref{fig:pfaf}. The $v_{b}(d)$ and $v_{f}(d)$ curves in Fig.~\ref{fig:velocity} are smooth and exhibit very similar dependences on $d$, implying that our method of calculating the velocity is reasonably valid. For the edge-reconstructed phase, where multiple edges are supposed to exist, the velocities calculated here are those of the first (innermost) edge.

An important shortcoming of the non-interacting model that we have used so far ought to be noted. When the dispersion goes negative, it becomes possible to construct, within this model, states with lower and lower energies at larger and larger wave vectors. This clearly does not happen in a realistic system, because the confinement potential introduces a significant energy cost to the creation of such modes. This problem can be addressed by introducing interactions between the effective particles \cite{PhysRevLett.91.036802}. We have found in our studies that for wave vectors up to the minimum in the dispersion, the non-interacting model is reasonably accurate, but it is less accurate for larger wave vectors, which we interpret as a signature of such interactions. We cannot exclude the possibility that for larger systems, the minimum energy state will contain many bosons and fermions (recall that the fermion number must be even). We have not explored this issue further. 

\subsection{Model II for the 5/2 edge}

\begin{figure}
\hspace{-3mm}\resizebox{0.4\textwidth}{!}{\includegraphics{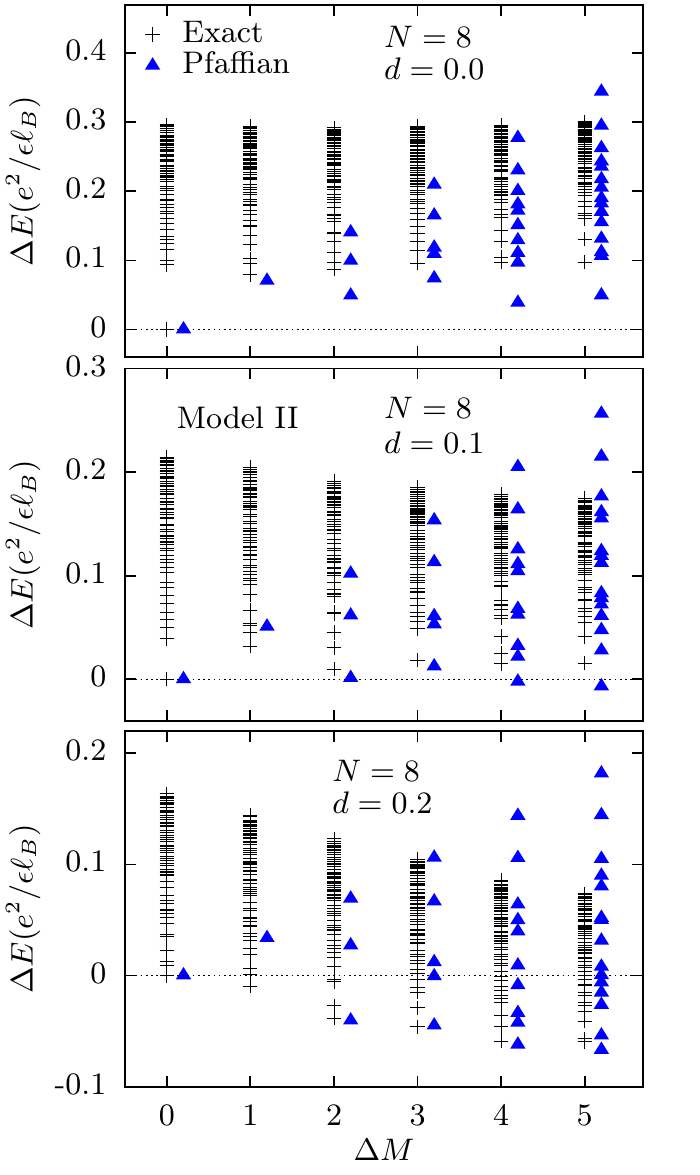}}\\
\vspace{-4mm}
\caption{(Color online) Comparison of the edge spectra obtained by diagonalizing the Model II Coulomb interaction within the entire Hilbert space (black pluses) and the Pfaffian basis (blue triangles) for $N=8$ particles. The energies are measured relative to the ground state energy in both spectra. Similar overall trends are also seen for the lowest spectral branch as in Model I.}
\label{fig:exact2}
\end{figure} 

In all the above calculations, we have neglected the electrons in the completely filled LLL. We now consider how they influence the results. For a $5/2$ state with $N$ electrons in the 2LL at filling factor $\nu_{0}=1/2$ on a disk, we model the LLL as a background with $4N$ static electrons uniformly distributed on the disk. The total amount of positive charges in the system is $5N$, and they are placed on a neutralizing disk at a setback distance $d$ from the electron disk. 
The disks corresponding to the positive background and the lowest filled LL have the same radii $R=\sqrt{2N/\nu_{0}}\ell_B$.  The Coulomb interaction in Model II still consists of three terms as in the first line of Eq.~(\ref{eq1}): $V_{\rm ee}$ is the same as in Model I; $V_{\rm eb}$ is the Coulomb interaction between the 2nd LL electrons and the two background charged disks (one positive and one negative); $V_{\rm bb}$ includes the self Coulomb energy of both the positive charged disk and negative charged disk as well as the interaction between the two background disks.

\begin{figure*}
\resizebox{0.84\textwidth}{!}{\includegraphics{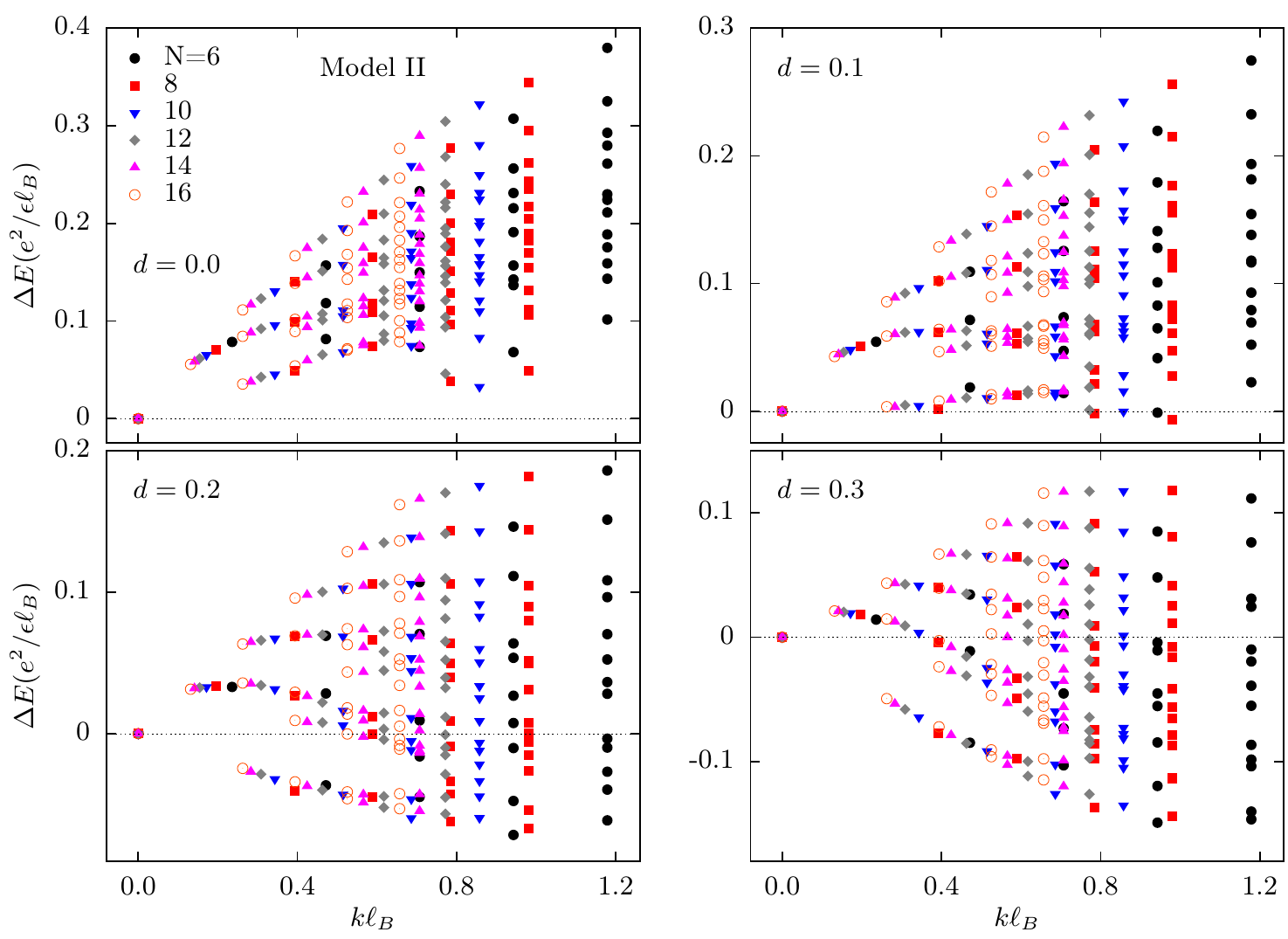}}\\
\vspace{-4mm}
\caption{(Color online) Edge spectra of  $5/2$ state as a function of the physical momentum k for $N=6-16$ particles. The energies are obtained by diagonalizing the full Hamiltonian of Model II within the Pfaffian edge basis. Data from N=8-16 collapse for the lowest spectral branch. Edge reconstruction occurs when $d \geq 0.1\ell_{B}$.}
\label{fig:52modeltwo}
\end{figure*}

We first study an $N=8$ system by diagonalizing the Hamiltonian in Model II within the full Hilbert space and the truncated Pfaffian subspace. As shown in Fig.~\ref{fig:exact2}, the lowest branches in the two energy spectra have similar overall trends for different values of $d$. This demonstrates that the approximation of using the truncated Pfaffian subspace is still valid for Model II. Fig.~\ref{fig:52modeltwo} shows the energy spectra within the Pfaffian subspace of different systems (up to $N=16$) using Model II. Data collapse is achieved as in Model I. Edge reconstruction occurs more easily than in Model I (see Fig.~\ref{fig:pfaf}), with the critical setback distance being only $0.1\ell_B$. This is to be expected since the LLL electrons are on the same plane as the 2LL electrons. The repulsion between the 2LL electrons and the LLL electrons is stronger on average than the attraction between the 2LL electrons with the positively charged disk at setback distance $d$, thereby weakening the confinement potential.

\begin{figure}
\hspace{-5mm}\resizebox{0.41\textwidth}{!}{\includegraphics{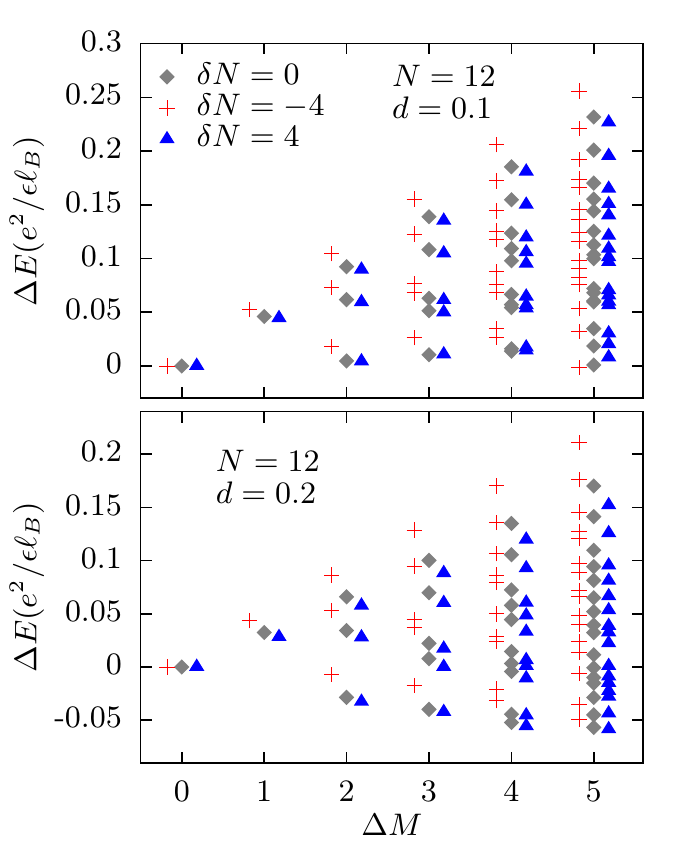}}\\
\vspace{-4mm}
\caption{(Color online) The edge spectra for $5/2$ systems with $N=12$ particles and $4N+\delta N$ electrons in LLL. The variation of the electron number in the LLL does not have appreciable influence on edge reconstruction.}
\label{fig:deltaN}
\end{figure}

In realistic physical systems, the edge of the LLL may not coincide with that of the 2LL, {\em i.e.}, the sizes of the two electrons disks could be different and the number of electrons in the LLL may not be $4N$. To understand how the number of electrons in the LLL affects edge reconstruction, we consider a $\nu=5/2$ state with $N=12$ particles in the 2LL and vary the number of electrons in the LLL while keeping the density $\rho$ fixed. To be specific, we assume there are $N_{1}=4N+\delta N$ electrons in the LLL with a radius $R_{1}=\sqrt{2N_{1}/\nu_{1}}\ell_{B}$ and $\nu_{1}=2$, while the positively charged background has a radius $R_{2}=\sqrt{2N_{2}/\nu_{2}}\ell_{B}$ with $N_{2}=5N+\delta N$ and $\nu_{2}=5/2$. Fig.~\ref{fig:deltaN} shows the energy spectra corresponding to $\delta N=-4$, $\delta N=0$, and $\delta N=4$. These three systems have similar features and they all have critical edge reconstruction distances $d_{c} \sim 0.1\ell_{B}$. This suggests that the critical setback distance is not particularly sensitive to the details of the relative occupations of the various LLs.

\section{Effective approach for reconstructed edge and experimental consequences \label{sec:exp}}

In the previous section, we see that edge reconstruction in the Pfaffian state starts to occur at relatively small values of $d$.  Therefore, the Pfaffian state is very likely to be edge-reconstructed in its experimental realization. This in general sense makes the edge physics non-universal. However, the effective theory description of the edge states described below indicates the presence of other fixed points for this system, which are described, in general, by different scaling exponents. The results of this section are summarized in subsection \ref{summary} and a reader who is not interested in the technical details but only in the final results can directly go there.

Let us start by first introducing the formalism that will be used in what follows. For the derivation of the formulas, see Ref. \onlinecite{PhysRevB.57.10138}. Let us consider an edge theory whose bosonic sector is described by 
\bea
S_{b} = \frac{1}{4 \pi} \int d\tau\,dx \left(K_{ij} \, \partial_{\tau} \phi_i \, \partial_x \phi_j + V_{ij}  \, \partial_{x} \phi_i \, \partial_x \phi_j \right), \label{eq:action}
\eea
where $i, j = 1, \cdots, n$; $n$ is the number of edge modes; $K$ is a symmetric integer matrix; and $V$ is a symmetric positive matrix. The filling factor is given by $\nu = t^T \cdot K^{-1} \cdot t$,
where the vector $t$ specifies the charges of quasiparticles. 

Let us consider an operator given by ${\cal O}_{\ell} = e^{i \ell_i \phi_i}$. Its charge is given by $q_{\ell} = t^T \cdot K^{-1} \cdot \ell$ and its exchange statistics with respect to another operator ${\cal O}_k$ (which can be itself) is given by $\theta_{k \ell} = \pi \, k^T \cdot K^{-1} \cdot \ell$.

In order to determine the Hall conductivity and the scaling dimension of operator ${\cal O}_{\ell}$, we need to diagonalize the action in Eq. (\ref{eq:action}). First, let us consider a basis transformation $\phi' = M_1^{-1} \cdot \phi$, under which 
\be
K' = M_1^T \cdot K \cdot M_1 =    \begin{pmatrix} 
      -\mathbb{I}_{n_-} & 0 \\
      0 & \mathbb{I}_{n_+}  \\
   \end{pmatrix},
\ee
where $\mathbb{I}_{n_{\pm}}$ is an $n_{\pm} \times n_\pm$ identity matrix and $n_- + n_+ = n$. Next, we can diagonalize $V'=M_1^T \cdot V \cdot M_1 $ by
\be
V'' = M_2^T \cdot M_1^T \cdot V \cdot M_1 \cdot M_2,
\ee
where $V''$ is a diagonal matrix and $M_2 \in SO(n_-,n_+)$ such that $K'' = K'$. We can express the second basis transformation as $M_2 = B \cdot R$, where $R$ is an orthogonal matrix, \textit{i.e.}, the rotation, and $B$ is a positive matrix, \textit{i.e.}, the pure boost of Lorentz group. It turns out that
\be
\sigma_H = 2 \, {t'}^T \cdot \Delta  \cdot {t'},
\ee
where
\be
\Delta = \frac{1}{2} \, M_1 \cdot B^2 \cdot M_1^T.
\ee
Furthermore, the scaling dimension of an operator ${\cal O}_{\ell'}$ is given by 
\be
\Delta_{\ell'} = {\ell'}^T \cdot \Delta \cdot {\ell'}.
\ee

For the Pfaffian state, along with the bosonic sector, there is also a Majorana fermion sector, which is described by chiral Ising conformal field theory (CFT). The primary field operators of the chiral Ising CFT are the identity $\mathds{1}$, the Majorana fermion $\psi$ and the spin operator $\sigma$. The Ising CFT sector is neutral and commutes with the bosonic sector. Its statistics is 
\be
\theta_{\psi \psi} = \pm \pi, \qquad \theta_{\psi \sigma} = \pm \pi/2,
\ee
where the positive (negative) sign corresponds to the backward (forward) moving modes. One can then construct electron operators by looking for charge 1 fermionic operators. The quasiparticles are then operators that are local with respect to all the electron operators, \textit{i.e.}, the phases induced by moving a quasiparticle around any electron operators are integer multiples of $2\pi$.

In previous study \cite{overbosch2008phase}, it was assumed that the edge reconstruction is driven by an instability in the bosonic sector such that the edge reconstruction results in two additional bosons but without any additional Majorana fermions. However, as we have seen in the previous section, we found that edge reconstruction in our models is driven by an instability in the Majorana fermion sector. In the following, we review the case of boson-driven edge reconstruction and then describe the edge theory of fermion-driven edge reconstruction.

\subsection{Boson-driven edge reconstruction}

\begin{figure}[h] 
   \centering
   \includegraphics[width=3in]{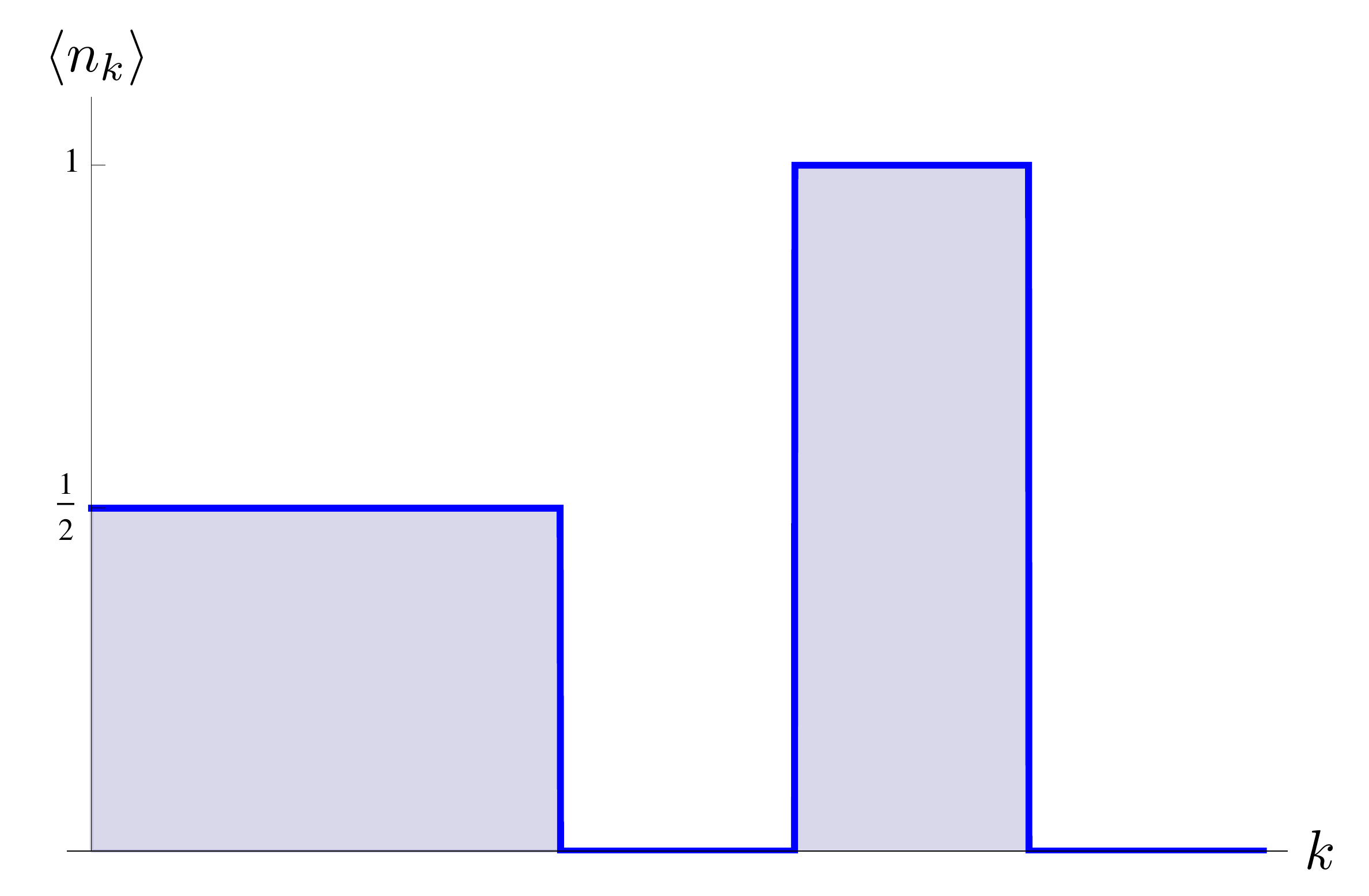} 
   \caption{Momentum occupation distribution for boson-driven edge reconstruction.}
   \label{fig:boson}
\end{figure}

In this case \cite{overbosch2008phase}, the momentum occupation distribution (within Hartree-Fock approximation) is assumed to be as depicted in Fig. \ref{fig:boson}. There is only one Majorana mode and the bosonic sector of theory is described by
\be
K =    \begin{pmatrix} 
      -1 & 0 & 0 \\
      0 & 1 & 0 \\
      0 & 0 & 2
   \end{pmatrix},
\qquad t =    \begin{pmatrix} 
      1  \\
      1  \\1  \\
   \end{pmatrix},
\ee
or equivalently, 
\be
K =    \begin{pmatrix} 
      -1 & 0 & 0 \\
      0 & 1 & 0 \\
      0 & 0 & 2
   \end{pmatrix},
\qquad t =    \begin{pmatrix} 
      0  \\
      0  \\1  \\
   \end{pmatrix}.
\ee
In the latter basis, in which only one of the quasiparticles is charged, we find that
\be
M_1= \begin{pmatrix} 
      1 & 0 & 0 \\
      0 & 1 & 0 \\
      0 & 0 & \frac{1}{\sqrt{2}}
   \end{pmatrix},
\ee
and parametrizing the boost such that
\be
B^2 = \begin{pmatrix} 
      \gamma & \beta_1 \gamma & \beta_2 \gamma \\
      \beta_1 \gamma & 1+ \frac{\beta_1^2 \gamma^2}{\gamma+1} & \frac{\beta_1 \beta_2 \gamma^2}{\gamma+1} \\
      \beta_2 \gamma & \frac{\beta_1 \beta_2 \gamma^2}{\gamma+1} & 1+ \frac{\beta_2^2 \gamma^2}{\gamma+1}   \end{pmatrix}, \label{eq:boost}
\ee
where $\gamma = 1/\sqrt{1-\beta^2}$, $\beta^2=\beta_1^2 + \beta_2^2$ and $|\beta| \leq 1$, yields
\be
\sigma_H = \frac{1}{2} \left(1+ \frac{\beta_2^2}{1-\beta^2 + \sqrt{1-\beta^2}}\right).
\ee
We note that when $\beta_2 \ne 0$, the charge mode is mixed with the backward moving neutral mode and the conductance is not universal. This is a typical behavior of edge theories with counter-propagating modes \cite{PhysRevLett.72.4129,PhysRevB.57.10138}.

If there is a relevant neutral operator with Bose-Einstein statistics $\theta=\pm2\pi$, disorder induced tunneling will cause the theory to flow to a strong coupling fixed point, the so-called Kane-Fischer-Polchinski (KFP) fixed point \cite{PhysRevLett.72.4129}. However, in the present case, there exist no such operators. Instead, there is a ``null" operator,  which is a neutral charge operator with equal left and right conformal dimensions. It is given by $\psi \exp[i (2\phi_1 + \phi_2 + 2 \phi_3)]$ with $\beta_1=\beta_2=0$. The bosonic part of the null operator $\exp[i (2\phi_1 + \phi_2 + 2 \phi_3)]$ is equivalent to a backward-moving complex fermion, which, in turn, is equivalent to two backward moving Majorana fermions. The bosonic part of the null operator couples to forward moving Majorana, resulting in gapping two of the Majoranas but  leaving a gapless backward-moving Majorana fermion. This is the so-called Majorana-gapped phase \cite{overbosch2008phase}. 

At this new phase, the Lagrangian density is given by
\begin{eqnarray}
{\cal L} &=& \frac{1}{4 \pi} \left[\partial_x \phi_c \left(\partial_\tau + v_c \partial_x\right) \phi_c + \partial_x \phi_n \left(\partial_\tau + v_n \partial_x\right) \phi_n \right. \nonumber\\
& &   \left. \qquad \quad +\, i \lambda \left(-\partial_\tau + v_{\lambda} \partial_x\right) \lambda \right],
\end{eqnarray}
with $t=(1/\sqrt{2},0)^T$. In this Majorana-gapped phase the charged bosonic mode $\phi_c$ and the neutral bosonic mode $\phi_n$ are decoupled, and therefore, $\sigma_H=5/2$ after we include the contribution from the LLL. 

The electron operators for the Majorana gapped phase of the edge-reconstructed Pfaffian state are given by $e^{i (\sqrt{2} \phi_c + \phi_n)}$ and $\lambda e^{i \sqrt{2} \phi_c}$, both with scaling dimension $\Delta = 3/2$. The most relevant quasiparticles in each sectors of Ising CFT along with their charge $q$ and scaling dimension $\Delta$ are:
\begin{eqnarray}
\mathds{1}-{\rm sector:}&&  \quad e^{i \phi_c/\sqrt{2}}, \quad q=1/2, \quad \Delta=1/4, \nonumber \\
\lambda-{\rm sector:}&&  \quad \lambda e^{i \phi_c/\sqrt{2}}, \quad q=1/2, \quad \Delta=3/4, \\
\sigma-{\rm sector:}&&  \quad \sigma e^{i \phi_c/(2\sqrt{2}) \pm \phi_n/2},  \quad q=1/4, \quad \Delta = 1/4. \nonumber 
\end{eqnarray}

\subsection{Majorana-driven edge reconstruction}

\begin{figure}[h] 
   \centering
   \includegraphics[width=3in]{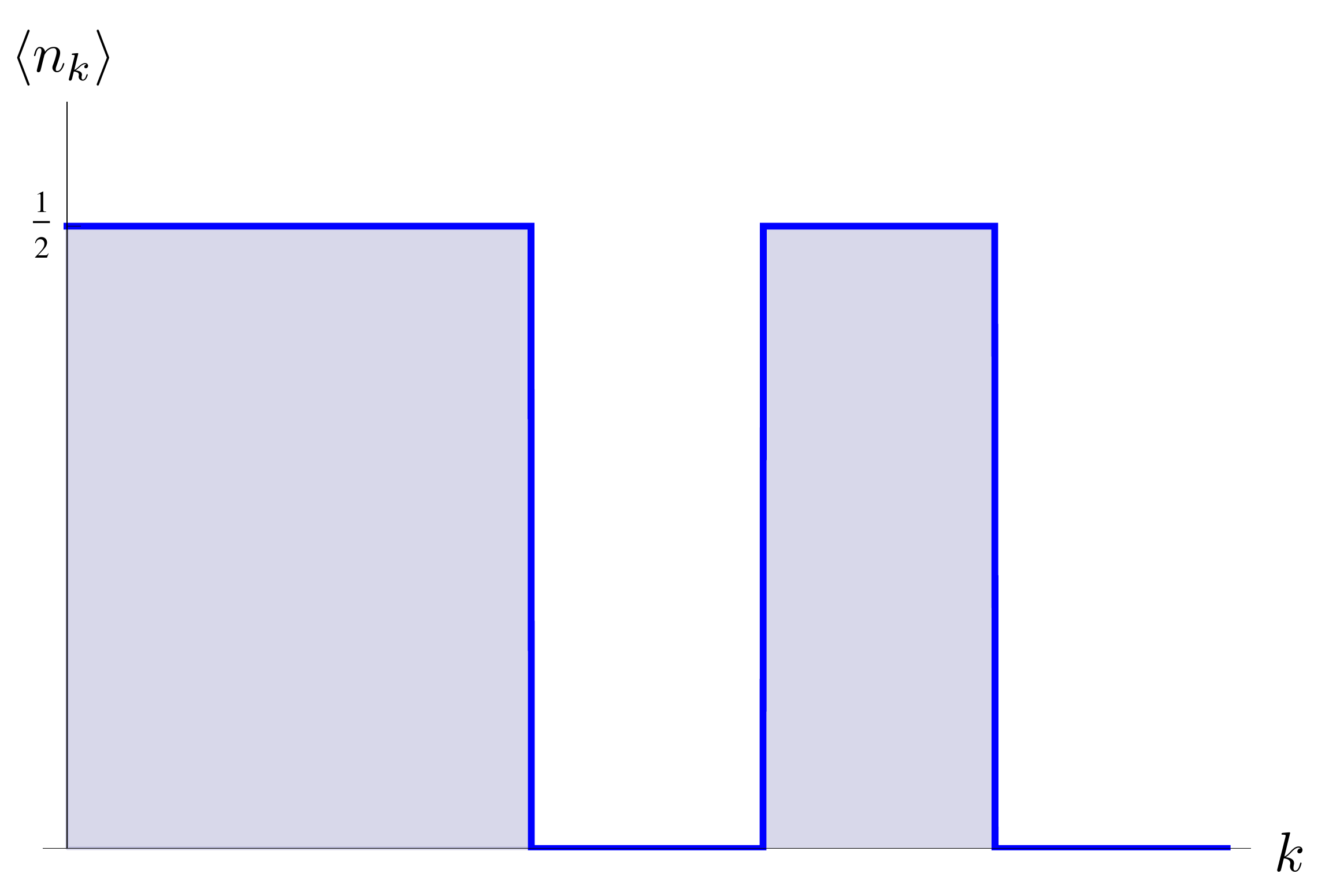} 
   \caption{Momentum occupation distribution for Majorana-driven edge reconstruction.}
   \label{fig:majorana}
\end{figure}

Here, we have two forward-moving and one backward-moving Majorana modes. However, the null operators obtained by combining the forward-moving modes with the backward-moving mode 
\be
S_m = i \int d\tau\,dx \left(m_{12}\, \psi_1 \psi_2 + m_{13} \,\psi_1 \psi_3 \right),
\ee
where $\psi_1$ is backward moving and $\psi_{2,3}$ are forward moving modes, are relevant and therefore gap $\psi_1$ and the linear combination $(m_{12} \,\psi_2 + m_{13} \,\psi_3)/\sqrt{m_{12}^2 + m_{13}^2}$, leaving $\psi \equiv (m_{12} \,\psi_2 - m_{13}\, \psi_3)/\sqrt{m_{12}^2 + m_{13}^2}$ gapless.

The bosonic sector is given by (see Fig. \ref{fig:majorana})
\be
K =    \begin{pmatrix} 
      -2 & 0 & 0 \\
      0 & 2 & 0 \\
      0 & 0 & 2
   \end{pmatrix},
\qquad t =    \begin{pmatrix} 
      1  \\
      1  \\1  \\
   \end{pmatrix},
\ee
or equivalently, 
\be
K =    \begin{pmatrix} 
      0 & -2 & 0 \\
      -2 & 0 & 0 \\
      0 & 0 & 2
   \end{pmatrix},
\qquad t =    \begin{pmatrix} 
      0  \\
      0  \\1  \\
   \end{pmatrix}.
\ee
As in the previous case, in the second basis, we find that
\be
M_1= \begin{pmatrix} 
      \frac{1}{2} & -\frac{1}{2} & 0 \\
      \frac{1}{2} & \frac{1}{2} & 0 \\
      0 & 0 & \frac{1}{\sqrt{2}}
   \end{pmatrix},
\ee
and using the boost of Eq. (\ref{eq:boost}) , we have
\be
\sigma_H = \frac{1}{2} \left(1+ \frac{\beta_2^2}{1-\beta^2 + \sqrt{1-\beta^2}}\right).
\ee
Again, $\beta_2 = 0$ is the charge-unmixed point where the charge mode is decoupled from the backward moving neutral mode. 

In order to obtain universal conductance, we look for neutral bosonic operators with $\theta = \pm 2 \pi$. There are four such operators (eight if we include their Hermitian conjugates) and they are given by
\bea
{\cal O}_1 &=& \exp[i(\phi_{n1} + 2 \phi_{n2})], \nonumber\\
&& \Delta_1 = \frac{(10 +6 \beta_1)(1+ \sqrt{1-\beta^2})-\beta_2^2}{8 (1-\beta^2 + \sqrt{1-\beta^2})}, \nonumber \\
{\cal O}_2 &=& \exp[i(2 \phi_{n1} + \phi_{n2})], \nonumber\\
&& \Delta_2 = \frac{ (10 -6 \beta_1)(1+ \sqrt{1-\beta^2})-\beta_2^2}{8 (1-\beta^2 + \sqrt{1-\beta^2})}, \nonumber \\
{\cal O}_3 &=& \exp[i(\phi_{n1} - 2 \phi_{n2})], \nonumber\\
&& \Delta_3 = \frac{(10 +6 \beta_1)(1+ \sqrt{1-\beta^2})-9\beta_2^2}{8 (1-\beta^2 + \sqrt{1-\beta^2})}, \nonumber \\
{\cal O}_4 &=& \exp[i(2 \phi_{n1} - \phi_{n2})], \nonumber\\
&& \Delta_4 = \frac{(10 -6 \beta_1)(1+ \sqrt{1-\beta^2})-9\beta_2^2}{8 (1-\beta^2 + \sqrt{1-\beta^2})},  \label{eq:tunops}
\eea
where $\phi_{n,i}$ are the two neutral modes in the $t=(0,0,1)^T$ basis. Using these operators, we can then add disorder induced tunneling terms into the action
\be
S_{\rm tunneling} = \sum_{i=1}^4 \int d\tau \, dx \, \left(\xi_i(x) \, {\cal O}_i + \xi_i^{\ast}(x) {\cal O}_i^{\dagger} \right),
\ee
where $\xi_i$'s are complex random variables and $\langle \xi(x) \xi^{\ast}(x') \rangle = D_i \delta(x-x')$, with $D_i$ the real-valued disorder strengths. The renormalization group (RG) equations up to linear order in the disorder strengths are given by
\bea
\frac{dD_i}{dl}= \left(3-2\Delta_i\right)D_i, \label{eq:RGD}\\
\frac{d\Delta_i}{dl} \propto -\left(\Delta_i^2-1\right)D_i. \label{eq:RGDelta}
\eea
Eq. (\ref{eq:RGD}) implies that these terms are relevant when $\Delta_i < 3/2$ while Eq. (\ref{eq:RGDelta}) implies that they are maximally relevant when $\Delta_i = 1$. If any of these terms is relevant, the theory will flow toward the strong disorder regime. Furthermore, if there is a point where all the relevant tunneling operators become maximally relevant, that point will be a (strong coupling) fixed point, \textit{i.e.}, the KFP fixed point.

\begin{figure}[htbp] 
   \centering
   \includegraphics[width=3in]{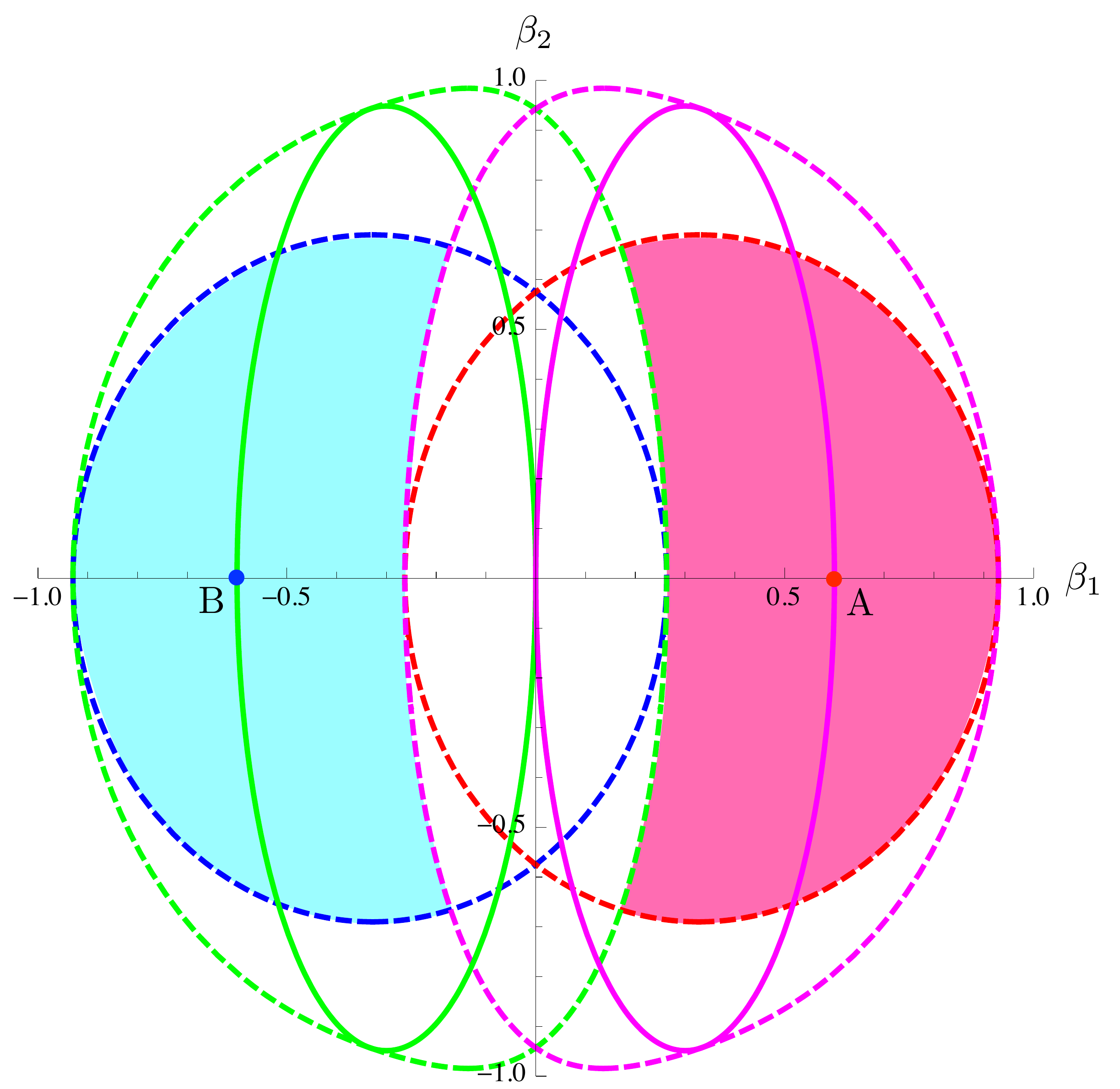} 
   \caption{Parameter regimes in which the tunneling terms are at least marginal, where the blue, red, green and magenta lines correspond to ${\cal O}_i$, $i=1, 2, 3$ and 4, respectively. Here, the dashed lines indicate where the operators become marginal ($\Delta = 3/2$) and the solid lines indicate where the operators ${\cal O}_{3,4}$ become maximally relevant ($\Delta=1$). The points A and B are the points where the operators ${\cal O}_{1,2}$ are maximally relevant, which also happen to be the KFP fixed points.}
   \label{fig:dimension}
\end{figure}

The parameter regimes in which the operators in Eq. (\ref{eq:tunops}) are relevant are depicted in Fig. \ref{fig:dimension}. The situation is clear when the interaction corresponds to a point within the shaded regions. Within the right shaded region, ${\cal O}_{1,3}$ are irrelevant while ${\cal O}_{2,4}$ are relevant, which will then drive the system to a KFP fixed point A with $(\beta_1,\beta_2) = (3/5,0)$. Within the left one, then ${\cal O}_{2,4}$ are irrelevant, while ${\cal O}_{1,3}$ will drive the system to a KFP fixed point B where $(\beta_1,\beta_2) = (-3/5,0)$. We note that at both of these fixed points, the Hall conductance has the expected value. For other regimes, we need to include higher order corrections in disorder strengths and such calculations are beyond the present work.

Let us now look at the scaling exponents at both fixed points. First of all, we have three linearly independent electron operators, given by
\bea
{\cal O}_{\rm el 1} &=& \psi \exp[i \,2 \phi_c], \nonumber \\
{\cal O}_{\rm el 2} &=& \psi \exp[i  (\phi_{n1} +2 \phi_c)], \nonumber \\
{\cal O}_{\rm el 3} &=& \psi \exp[i  ( \phi_{n2} + 2\phi_c)],
\eea
where $\phi_c$ is the charged mode. At fixed point A, their scaling dimensions are $3/2$, $13/8$ and $2$, respectively, while at fixed point B, the scaling dimensions are $3/2$, $2$ and $13/8$, respectively. 

The most relevant quasiparticle contents of fixed points A and B are given by
\begin{eqnarray}
\mathds{1}-{\rm sector:}&&  \quad e^{i \phi_c}, \quad q=1/2, \quad \Delta=1/4, \nonumber \\
\psi-{\rm sector:}&&  \quad \psi e^{i \phi_c}, \quad q=1/2, \quad \Delta=3/4, \\
\sigma-{\rm sector:}&&  \quad \sigma e^{i \phi_c/2},  \quad q=1/4, \quad \Delta = 1/8. \nonumber 
\end{eqnarray}

In the case of Majorana-driven edge reconstruction, along with $\psi_1 \psi_2$ and $\psi_1 \psi_3$, there are also several of null operators that contain the bosons. They are given by ${\cal O}^{\rm null}_1 =\psi_1 \exp[i (\phi_{n2}-\phi_{n1})]$ for $\beta_1=0$ and ${\cal O}^{\rm null}_{2,3} =\psi_{2,3} \exp[i (\phi_{n1}+\phi_{n2})]$ for $\beta_1=\beta_2=0$. When $\beta_1=0$ and $\beta_2 \ne 0$, only ${\cal O}^{\rm null}_1$ is turned on and we have
\bea
{\cal L} &=& \frac{1}{4 \pi} \left[\partial_x \phi_c \left(\partial_\tau + v_c \partial_x\right) \phi_c + \partial_x \phi_n \left(-\partial_\tau + v_n \partial_x\right) \phi_n \right. \nonumber\\
& & \qquad \quad + \, v_m \partial_x \phi_c \partial_x \phi_n + i \psi \left(\partial_\tau + v_{\psi} \partial_x\right) \psi \nonumber \\
& &   \left. \qquad \qquad +\, i \lambda_i \left(\partial_\tau + v_{\lambda_i} \partial_x\right) \lambda_i \right],
\end{eqnarray}
with $t = (1/\sqrt{2},0)^T$ and $i=1,2$. We note that since the charged mode and the backward-moving neutral mode are not decoupled, the Hall conductivity is not universal. There is a neutral bosonic operator $\exp(i \sqrt{2} \phi_n)$ that will drive the system to flow to a KFP fixed point where $v_m=0$. At this point, however, there are also null operators given by $\psi \exp(i \sqrt{2} \phi_n)$ (which is a linear combination of ${\cal O}^{\rm null}_2$ and ${\cal O}^{\rm null}_3$) and $\lambda_i \exp(i \sqrt{2} \phi_n)$. These gap $\phi_n$ and a linear combination of $\psi$ and $\lambda_i$, leaving three gapless Majorana modes: two forward and one backward moving. This is exactly what we would have obtained if we had turned on all ${\cal O}^{\rm null}_i$'s from the beginning, which is not surprising as $v_m=0$ if and only if $\beta_2=0$.

The remaining Majorana modes can then be gapped further, which in the end, leaves us with only one gapless forward moving Majorana fermion. Since the neutral bosonic modes have also been gapped, we recognize that this state is nothing but the original edge unreconstructed Pfaffian state. In other words, in the case of fermion-driven edge reconstruction, even though the null operators do not look trivial, their effect is only to drive the system back toward the edge unreconstructed state.

\subsection{Summary of results}
\label{summary}

\begin{table}[h]
\begin{ruledtabular}
\begin{tabular}{ l  c  c }
   & \quad $g_q$ \quad &\quad $e^{\ast}$ \qquad\\
  \hline
  Pfaffian without edge reconstruction &  1/4 & 1/4 \\
  \hline
  Bosonic edge reconstruction &  1/2 & 1/4 \\
  \hline
  Fermionic edge reconstruction & 1/4 & 1/4\\
\end{tabular}
\end{ruledtabular}
\caption{Tunnelling exponents and charges of quasiparticles for different candidates of $\nu =5/2$ FQH state.} \label{tab:gandq}
\end{table}

Since the Pfafian state consists of bosonic and Majorana fermionic edge modes, its edge reconstruction can be driven by an instability either in the bosonic sector, as assumed in the previous study \cite{overbosch2008phase}, or by an instability in the Majorana fermion sector, as we have found in Section \ref{sec:5/2}.

We have found that the electron tunneling exponent (at large time scale), as measured from the I-V characteristics for tunneling of an electron from an external Fermi liquid into the FQH edge, is the same for both bosonic and fermionic edge reconstructions. However, the quasiparticle exponent, relevant for tunneling between two edges of a single FQH system, can distinguish between the two reconstructions. In Table. \ref{tab:gandq}, we list quasiparticle characteristics of Pfaffian state with different edge reconstructions. There, $g_q$ is given by twice the scaling dimension of the most relevant quasiparticle operator and $e^{\ast}$ is the charge. For example, the low-temperature conductance of quasiparticle tunneling through a quantum point contact will be given by $G(T) \sim T^{2(q_q - 1)}$, where $T$ is the temperature \footnote{There can be extra complications in the tunneling process, see for example, Refs. \onlinecite{Carrega11} and \onlinecite{Carrega12}.}. We note that the $(g_q,e^{\ast})$-values of fermion-driven edge reconstructed Pfaffian are identical to that of the original Pfaffian state, while the $(g_q,e^{\ast})$-value of the boson-driven edge reconstructed Pfaffian is identical to that of anti-Pfaffian state. 

One can further distinguish these states by probing the neutral sector, some properties of which are summarized in Table \ref{tab:neutral}. The thermal Hall conductivity does not distinguish between the states. To do so, one needs to directly detect the neutral bosonic backwards moving mode akin to what was done in Ref. \onlinecite{bid2010observation}.

\begin{table}[h]
\begin{ruledtabular}
\begin{tabular}{ l | l  l  c}
  &  boson & fermion & $\kappa_H$ \qquad \\
  \hline
  Pfaffian &  none & forward & 3/2 \\
  \hline
  Bos. edge recon. &  forward & backward & 3/2\\
  \hline
  Ferm. edge recon. & anti-parallel & forward & 3/2\\
\end{tabular}
\end{ruledtabular}
\caption{The neutral sector for different candidates of $\nu =5/2$ FQH state along with their thermal Hall conductance $\kappa_H$.} \label{tab:neutral}
\end{table}

\section{Edge spectra and reconstruction at $\nu=7/3$ \label{sec:7/3}}

As discussed in the introduction, the physics of the edge states at $\nu=7/3$ is also of relevance to several experiments. In this section, we study the possibility of edge reconstruction at $7/3$ within Model I by diagonalizing the 2LL Coulomb interaction in the truncated subspace given by the CF trial wave functions.

\begin{figure}
\hspace{-5mm}\resizebox{0.42\textwidth}{!}{\includegraphics{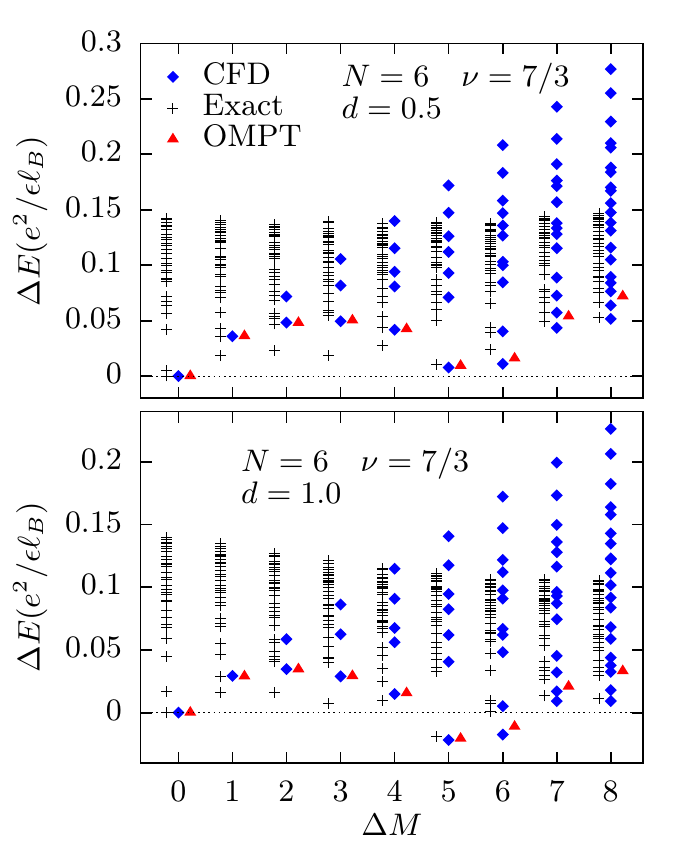}}\\
\vspace{-4mm}
\caption{(Color online) The edge excitation spectra obtained by diagonalizing the full Hamiltonian within the full basis (black pluses) and CF basis (blue diamonds), and the energy dispersion of OMPT wave function (red triangles), for $\nu=7/3$ system with $N=6$ at $d=0.5\ell_{B}, 1.0\ell_{B}$. The OMPT dispersion matches very well with the states that are identified as single-boson edge excitations in CF spectrum.}
\label{fig:73compare}
\end{figure}

The wave functions used here are not very accurate because the screening by CF excitons causes a strong renormalization in the $\nu=7/3$ state~\cite{Balram13b}. Nevertheless, we will compute the energy spectra within the framework of Model I using LLL wave functions together with the effective interaction to mimic the physics of the 2LL. To test the validity of this approach, we first compare the results obtained in exact diagonalization and the CF diagonalization for a $N=6$ system as shown in Fig.~\ref{fig:73compare}. Unlike the edge excitations at $1/3$~\cite{Jolad:2010}, the CF basis states are not very accurate approximations of the exact eigenstates at $7/3$, but they do capture the qualitative behavior of the edge reconstruction. (We note that the edge boson in exact diagonalization has a non-monotonic dispersion, as indicated by an additional ``minimum" in its dispersion at $\Delta M=3$ in Fig.~\ref{fig:73compare}. The presence of this additional structure does not affect the edge reconstruction physics, however, and therefore we have not further pursued its physical origin.) We have also calculated the dispersion relation using the OMPT method and found that it captures the lowest energy branch of the CF energy spectra for $\Delta M<6$, which confirms that these are the single-boson edge excitations.  For larger $\Delta M$, the single-boson excitations no longer have the lowest energy. We find that the OMPT trial wave functions also successfully reproduce the single-boson mode of the CF energy spectra at $\nu=7/3$, analogously to what was found for the $5/2$ state.  

\begin{figure*}
\resizebox{0.87\textwidth}{!}{\includegraphics{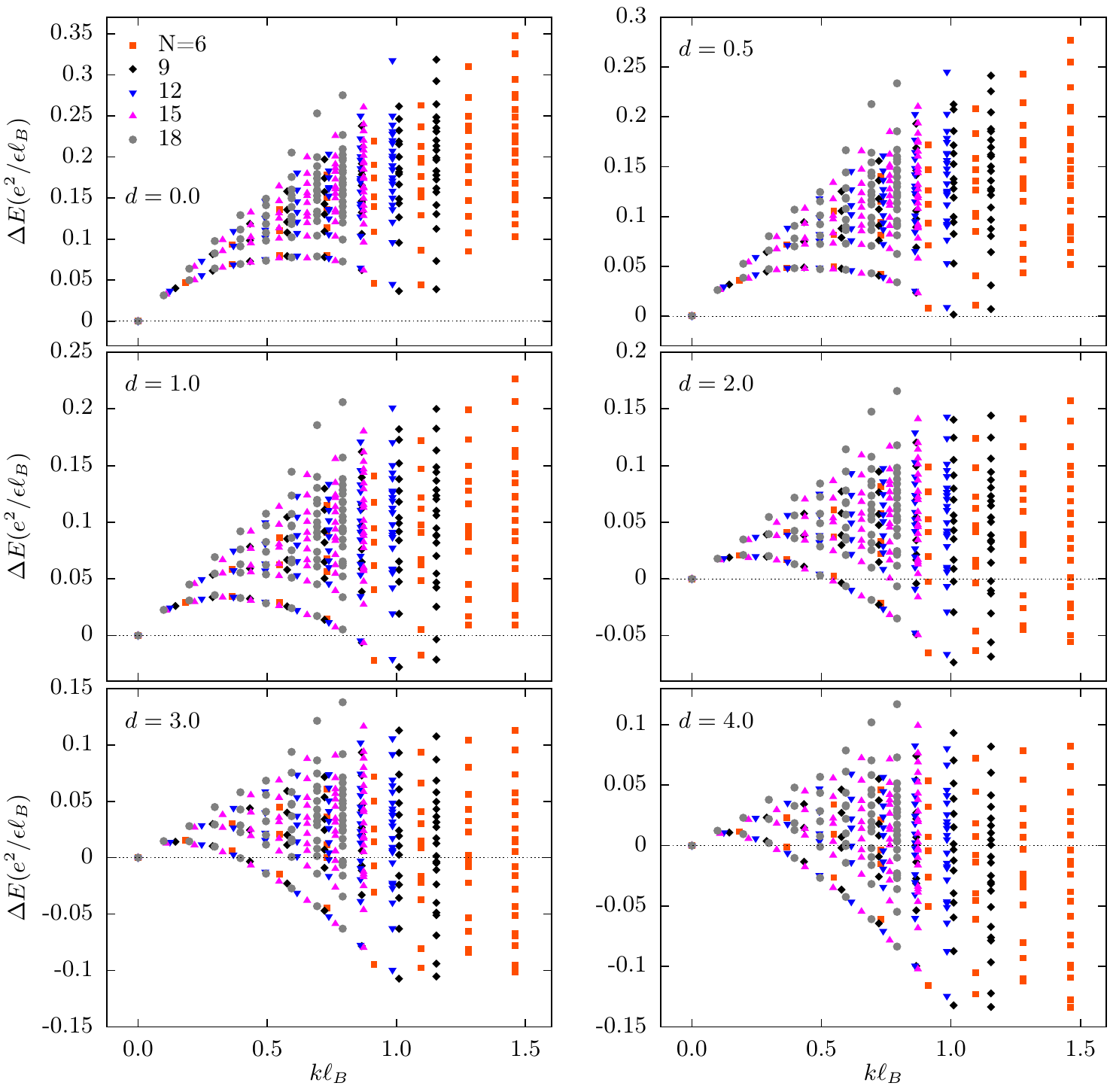}} \\
\vspace{-4mm}
\caption{(Color online) Spectrum of edge excitations at $7/3$ as a function of the wave vector $k$ for $N=6-18$ particles. The energies are obtained within a restricted CF basis, as explained in the main text. Data collapse can be seen for the lowest spectral branch. Edge reconstruction occurs when $d\geq0.5l$.}
\label{fig:spectra73}
\end{figure*}

To approach thermodynamic limit, we use the following scaling relation between the physical momentum $\delta k$ and the relative angular momentum $\Delta M$
\begin{equation}
\delta k=\frac{\Delta M}{\sqrt{6(N-1)}\ell_B}.
\end{equation}
We then plot the CF energy spectra versus $k$ in Fig.~\ref{fig:spectra73} for $N=6-18$, $d=0-4.0\ell_B$, and $\Delta M=0-8$. Data collapse of different systems can be seen in the branches with lowest and second lowest energies. The critical distance for edge reconstruction is around $d_{c}=0.5\ell_B$, which is significantly small compared to the $d_{c}\approx1.5\ell_B$ of the $1/3$ FQH state found in previous works~\cite{Wan02,Jolad:2010}. Including the lowest filled Landau level is likely to further reduce the critical value of $d$ for the same reasons as those explained in the context of 5/2. Our results indicate that edge reconstruction occurs more easily in the 2LL than in the LLL.

\section{Discussions and Conclusions \label{sec:d&c}}

We have performed an exhaustive study of the possibility of edge reconstruction of the $\nu=5/2$ and the $\nu=7/3$ FQH states, with a quasi-realistic treatment of the background neutralizing charge located in a layer at a distance $d$. We find that edge reconstruction occurs more readily in the second LL than in the lowest LL, and that the edges of both the 5/2 and 7/3 FQH states are reconstructed in current experiments. This physics should therefore be included in the analyses of the various experiments that attempt to probe the nature of the bulk FQH state through the properties of its edges. 

It is natural to ask how edge reconstruction affects various experimentally measurable quantities. The immediate consequence is the loss of topological properties of the FQH edge. However, in certain idealized limits the effective bosonic field theory approach suggests flows to new fixed points characterized with different exponents. Using the effective edge theory approach, we calculate the $(g_q,e^{\ast})$-values of quasiparticles that will dominate tunneling experiments. We saw that Majorana-driven edge reconstructed Pfaffian state is very similar to the original Pfaffian state in this regard, while boson-driven edge reconstructed state is similar to anti-Pfaffian state. To further distinguish these states, one needs to probe the neutral sector by directly probing the counter propagating modes \cite{bid2010observation}.

We stress that the conclusions presented above are based on several assumptions, listed in the introduction an elsewhere. Specifically, we have uncritically assumed the quantitative validity of the Pfaffian model for the edge excitations at 5/2, the CF model for the edge excitations at 7/3, and the effective theory for deducing the transport properties of the edge. Our results are also based on numerical studies of finite systems. The validity of our conclusions is contingent upon the validity of these assumptions. 

\section*{Acknowledgements}

We thank Prof. Kun Yang for useful discussions and for insightful comments on the manuscript. This work is supported by DOE under Grant No. DE-SC0005042. High-performance computing resources and services are provided by Research Computing and Cyberinfrastructure, a unit of Information Technology Services at The Pennsylvania State University. 

\appendix

\section{Effective electron-background interaction \label{appx:eb}}

\begin{figure*}
\resizebox{\textwidth}{!}{ \includegraphics{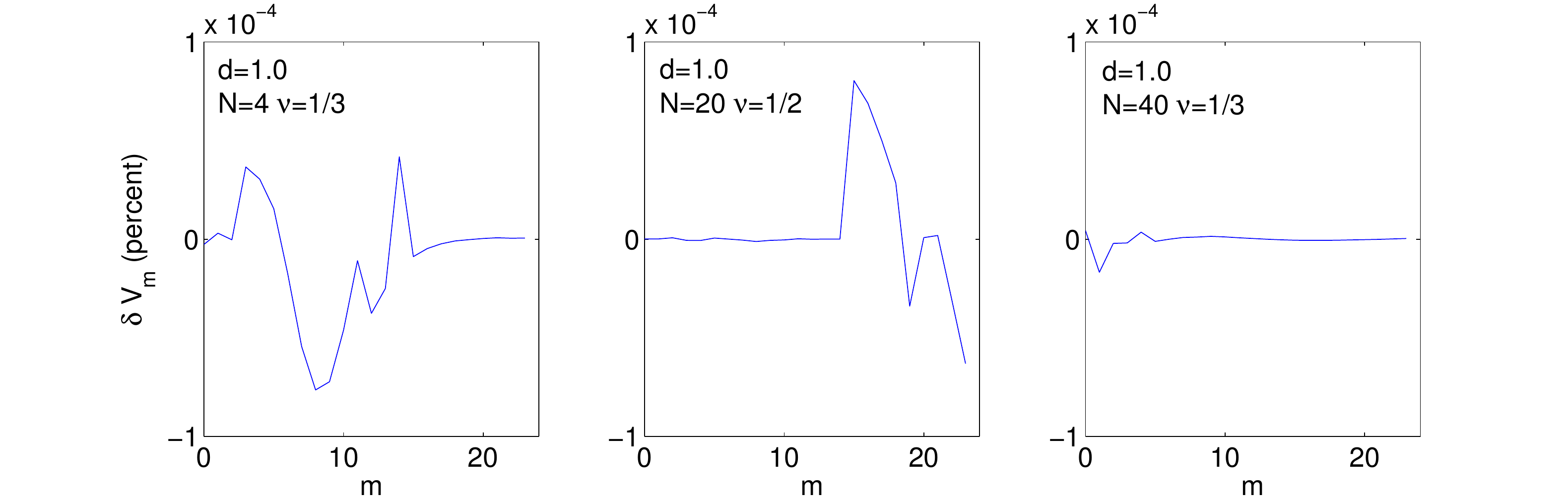}}\\
\vspace{-3mm}
\caption{The difference (in percent) between $n=0$ LL matrix elements $\langle m |V_{\rm eb}^{(\text{eff})}(|{\bf r}|)|m\rangle$ and $n=1$ LL matrix elements $\langle 1, m |V_{\rm eb}(|{\bf r}|)|1, m\rangle$ for different systems with $d=1.0$. Here, the $V_{\rm eb}^{(\text{eff})}$ takes the form of Eq. (\ref{eq:Vebeffform}) with $D_2=D_4=0, D_3=0.5, D_5=-1.5$. }
\label{fig:percent}
\end{figure*}

A method of identifying effective electron-electron interaction to mimic the 2LL Coulomb interaction was proposed in Ref.~\onlinecite{ChuntaiShi:2008}. Following the same line of thought, we develop an effective interaction for the electron-background interaction $V_{\rm eb}$.

The second quantized Hamiltonian in the LLL is 
\begin{eqnarray}
H = \frac{1}{2} \sum_{r,s,t,u} \langle r,s \left| V_{\rm ee} \right| t,u \rangle a_r^\dagger a_s^\dagger a_u a_t \nonumber \\
+ \sum_m \langle m \left| V_{\rm eb} \right| m \rangle a_m^\dagger a_m + V_{\rm bb}. 
\label{eq:2ndqH}
\end{eqnarray}
All real space electron-background interactions that lead to the same value of $\langle m\left|V_{\rm eb}\right|m\rangle$ are identical for the LLL problem. For electrons confined to the $n$-th Landau level, we need the matrix elements $\langle n, m\left|V_{\rm eb}\right|n, m\rangle$ where the LL eigenstates are given by 
\begin{equation}
|n,m\rangle\equiv \frac{(a^\dagger)^n}{\sqrt{n!}}\frac{(b^\dagger)^m}{\sqrt{m!}}|0\rangle, \label{eq:occ}
\end{equation}
We choose $m$ to be $0,1,\cdots$ for any LL; the angular momentum in this notation is given by $m-n=-n,-n+1,\cdots$. The problem of electrons with confinement energy $V_{\rm eb}(\mathbf{r})$ in the $n$-th LL for $n{\geq}1$ is mathematically equivalent to the problem of electrons in the LLL with an effective confinement energy $V_{\rm eb}^{(\text{eff})}(\mathbf{r})$ satisfying
\begin{equation}
\langle n,m \left| V_{\rm eb}(\mathbf{r}) \right| n,m \rangle = \left\langle m \left| V^{({\rm eff})}_{\rm eb}(\mathbf{r}) \right| m \right\rangle. \label{eq:Vebeff}
\end{equation}
Using Fourier transform
\begin{equation}
V_{\rm eb}(\mathbf{r}) = \int{\frac{d^2{\bf k}}{(2\pi)^2} V(\mathbf{k})e^{i\mathbf{k}\cdot\mathbf{r}}},
\end{equation}
and the identity
\begin{equation}
\langle n,m \left| e^{i\mathbf{k}\cdot\mathbf{r}} \right| n,m \rangle = L_n \left( \frac{\mathbf{k}^2}{2} \right) \langle m| e^{i\mathbf{k}\cdot\mathbf{r}} | m \rangle, \label{eq:id}
\end{equation}
we obtain the effective confinement energy:
\begin{equation}
V^{({\rm eff})}_{\rm eb}(\mathbf{k}) = L_n\left(\frac{\mathbf{k}^2}{2}\right) V(\mathbf{k}), 
\label{eq:Lag}
\end{equation}
where $L_n$ denotes the Laguerre polynomial.

The electron-background interaction $V_{\rm eb}$ can only be evaluated using numerical integration, so it seems difficult to obtain an analytic form for the effective interaction. To proceed, we write down the one electron component of $V_{\rm eb}(\mathbf{r})$ as
\begin{equation}
\begin{aligned}
V_{\rm eb}({\bf r}) = -\frac{\rho_0 e^2}{\epsilon} \int_{\Omega_N}{d^2 {\bf r^\prime}  \frac{1}{ R ({\bf r, r^\prime} ,d)}},
\end{aligned}
\end{equation}
where $R ({\bf r, r^\prime},d)=\sqrt{| {\bf r}-{\bf r^\prime} |^2+d^2}$ and the integral is over a 2D disk of radius $R_N=\sqrt{2N/\nu}$. For $d\neq 0$, we take the effective interaction to have the form
\begin{widetext}
\begin{equation}
V_{\rm eb}^{(\text{eff})}({\bf r})= -\frac{e^2 \rho_0}{\epsilon} \int_{\Omega_N}{d^2 {\bf r^\prime}  \left( \frac{1}{ R} + D_2 \frac{1}{ R^2}+D_3 \frac{1}{ R^3} +D_4 \frac{1}{R^4} +D_5 \frac{1}{R^5} \right)},
\label{eq:Vebeffform2}
\end{equation}
\end{widetext}
where $R=R ({\bf r, r^\prime},d)$ is the same as defined earlier. For the $n=1$ LL, the parameters $D_k$ can be determined by matching the matrix elements on the left and right hand sides of Eq.~(\ref{eq:Vebeff}) using the linear regression method. Each matrix element is in fact a three dimensional integral which can be calculated numerically. Interestingly, the values of the coefficients $D_k$ of different systems (which have different radius $R_N=\sqrt{2N/\nu}$) turn out to be approximately independent of $N$ and $\nu$ for a given $d$. The validity of this effective interaction is seen by noticing that the difference between the two matrix elements in Eq.~(\ref{eq:Vebeff}) is less than $0.0001\%$, as seen in Fig.~\ref{fig:percent} for three different systems. Table~\ref{tab:D} shows the values for $D_k$ for some other values of $d$. 

For $d=0$ we use the interaction given in Eq.~(\ref{deq0}). The difference between $\langle m |V_{\rm eb}^{(\text{eff})}(r)|m\rangle$ and $\langle 1, m |V_{\rm eb}(r)|1, m\rangle$ for this interaction is on the order of $0.001\%$.

\bibliographystyle{apsrev4-1}
\bibliography{References,biblio_fqhe}

\begin{thebibliography}{67}%
\makeatletter
\providecommand \@ifxundefined [1]{%
 \@ifx{#1\undefined}
}%
\providecommand \@ifnum [1]{%
 \ifnum #1\expandafter \@firstoftwo
 \else \expandafter \@secondoftwo
 \fi
}%
\providecommand \@ifx [1]{%
 \ifx #1\expandafter \@firstoftwo
 \else \expandafter \@secondoftwo
 \fi
}%
\providecommand \natexlab [1]{#1}%
\providecommand \enquote  [1]{``#1''}%
\providecommand \bibnamefont  [1]{#1}%
\providecommand \bibfnamefont [1]{#1}%
\providecommand \citenamefont [1]{#1}%
\providecommand \href@noop [0]{\@secondoftwo}%
\providecommand \href [0]{\begingroup \@sanitize@url \@href}%
\providecommand \@href[1]{\@@startlink{#1}\@@href}%
\providecommand \@@href[1]{\endgroup#1\@@endlink}%
\providecommand \@sanitize@url [0]{\catcode `\\12\catcode `\$12\catcode
  `\&12\catcode `\#12\catcode `\^12\catcode `\_12\catcode `\%12\relax}%
\providecommand \@@startlink[1]{}%
\providecommand \@@endlink[0]{}%
\providecommand \url  [0]{\begingroup\@sanitize@url \@url }%
\providecommand \@url [1]{\endgroup\@href {#1}{\urlprefix }}%
\providecommand \urlprefix  [0]{URL }%
\providecommand \Eprint [0]{\href }%
\providecommand \doibase [0]{http://dx.doi.org/}%
\providecommand \selectlanguage [0]{\@gobble}%
\providecommand \bibinfo  [0]{\@secondoftwo}%
\providecommand \bibfield  [0]{\@secondoftwo}%
\providecommand \translation [1]{[#1]}%
\providecommand \BibitemOpen [0]{}%
\providecommand \bibitemStop [0]{}%
\providecommand \bibitemNoStop [0]{.\EOS\space}%
\providecommand \EOS [0]{\spacefactor3000\relax}%
\providecommand \BibitemShut  [1]{\csname bibitem#1\endcsname}%
\let\auto@bib@innerbib\@empty
\bibitem [{\citenamefont {Tsui}\ \emph {et~al.}(1982)\citenamefont {Tsui},
  \citenamefont {Stormer},\ and\ \citenamefont {Gossard}}]{Tsui82}%
  \BibitemOpen
  \bibfield  {author} {\bibinfo {author} {\bibfnamefont {D.~C.}\ \bibnamefont
  {Tsui}}, \bibinfo {author} {\bibfnamefont {H.~L.}\ \bibnamefont {Stormer}}, \
  and\ \bibinfo {author} {\bibfnamefont {A.~C.}\ \bibnamefont {Gossard}},\
  }\href {\doibase 10.1103/PhysRevLett.48.1559} {\bibfield  {journal} {\bibinfo
   {journal} {Phys. Rev. Lett.}\ }\textbf {\bibinfo {volume} {48}},\ \bibinfo
  {pages} {1559} (\bibinfo {year} {1982})}\BibitemShut {NoStop}%
\bibitem [{\citenamefont {Willett}\ \emph {et~al.}(1987)\citenamefont
  {Willett}, \citenamefont {Eisenstein}, \citenamefont {St\"ormer},
  \citenamefont {Tsui}, \citenamefont {Gossard},\ and\ \citenamefont
  {English}}]{Willett87}%
  \BibitemOpen
  \bibfield  {author} {\bibinfo {author} {\bibfnamefont {R.}~\bibnamefont
  {Willett}}, \bibinfo {author} {\bibfnamefont {J.~P.}\ \bibnamefont
  {Eisenstein}}, \bibinfo {author} {\bibfnamefont {H.~L.}\ \bibnamefont
  {St\"ormer}}, \bibinfo {author} {\bibfnamefont {D.~C.}\ \bibnamefont {Tsui}},
  \bibinfo {author} {\bibfnamefont {A.~C.}\ \bibnamefont {Gossard}}, \ and\
  \bibinfo {author} {\bibfnamefont {J.~H.}\ \bibnamefont {English}},\ }\href
  {\doibase 10.1103/PhysRevLett.59.1776} {\bibfield  {journal} {\bibinfo
  {journal} {Phys. Rev. Lett.}\ }\textbf {\bibinfo {volume} {59}},\ \bibinfo
  {pages} {1776} (\bibinfo {year} {1987})}\BibitemShut {NoStop}%
\bibitem [{\citenamefont {Moore}\ and\ \citenamefont
  {Read}(1991)}]{Moore1991362}%
  \BibitemOpen
  \bibfield  {author} {\bibinfo {author} {\bibfnamefont {G.}~\bibnamefont
  {Moore}}\ and\ \bibinfo {author} {\bibfnamefont {N.}~\bibnamefont {Read}},\
  }\href {\doibase 10.1016/0550-3213(91)90407-O} {\bibfield  {journal}
  {\bibinfo  {journal} {Nucl. Phys. B}\ }\textbf {\bibinfo {volume} {360}},\
  \bibinfo {pages} {362 } (\bibinfo {year} {1991})}\BibitemShut {NoStop}%
\bibitem [{\citenamefont {Greiter}\ \emph {et~al.}(1991)\citenamefont
  {Greiter}, \citenamefont {Wen},\ and\ \citenamefont
  {Wilczek}}]{PhysRevLett.66.3205}%
  \BibitemOpen
  \bibfield  {author} {\bibinfo {author} {\bibfnamefont {M.}~\bibnamefont
  {Greiter}}, \bibinfo {author} {\bibfnamefont {X.-G.}\ \bibnamefont {Wen}}, \
  and\ \bibinfo {author} {\bibfnamefont {F.}~\bibnamefont {Wilczek}},\ }\href
  {\doibase 10.1103/PhysRevLett.66.3205} {\bibfield  {journal} {\bibinfo
  {journal} {Phys. Rev. Lett.}\ }\textbf {\bibinfo {volume} {66}},\ \bibinfo
  {pages} {3205} (\bibinfo {year} {1991})}\BibitemShut {NoStop}%
\bibitem [{\citenamefont {Jain}(1989)}]{Jain89}%
  \BibitemOpen
  \bibfield  {author} {\bibinfo {author} {\bibfnamefont {J.~K.}\ \bibnamefont
  {Jain}},\ }\href {\doibase 10.1103/PhysRevLett.63.199} {\bibfield  {journal}
  {\bibinfo  {journal} {Phys. Rev. Lett.}\ }\textbf {\bibinfo {volume} {63}},\
  \bibinfo {pages} {199} (\bibinfo {year} {1989})}\BibitemShut {NoStop}%
\bibitem [{\citenamefont {Lee}\ \emph {et~al.}(2007)\citenamefont {Lee},
  \citenamefont {Ryu}, \citenamefont {Nayak},\ and\ \citenamefont
  {Fisher}}]{Lee07}%
  \BibitemOpen
  \bibfield  {author} {\bibinfo {author} {\bibfnamefont {S.-S.}\ \bibnamefont
  {Lee}}, \bibinfo {author} {\bibfnamefont {S.}~\bibnamefont {Ryu}}, \bibinfo
  {author} {\bibfnamefont {C.}~\bibnamefont {Nayak}}, \ and\ \bibinfo {author}
  {\bibfnamefont {M.~P.~A.}\ \bibnamefont {Fisher}},\ }\href {\doibase
  10.1103/PhysRevLett.99.236807} {\bibfield  {journal} {\bibinfo  {journal}
  {Phys. Rev. Lett.}\ }\textbf {\bibinfo {volume} {99}},\ \bibinfo {pages}
  {236807} (\bibinfo {year} {2007})}\BibitemShut {NoStop}%
\bibitem [{\citenamefont {Levin}\ \emph {et~al.}(2007)\citenamefont {Levin},
  \citenamefont {Halperin},\ and\ \citenamefont {Rosenow}}]{Levin07}%
  \BibitemOpen
  \bibfield  {author} {\bibinfo {author} {\bibfnamefont {M.}~\bibnamefont
  {Levin}}, \bibinfo {author} {\bibfnamefont {B.~I.}\ \bibnamefont {Halperin}},
  \ and\ \bibinfo {author} {\bibfnamefont {B.}~\bibnamefont {Rosenow}},\ }\href
  {\doibase 10.1103/PhysRevLett.99.236806} {\bibfield  {journal} {\bibinfo
  {journal} {Phys. Rev. Lett.}\ }\textbf {\bibinfo {volume} {99}},\ \bibinfo
  {pages} {236806} (\bibinfo {year} {2007})}\BibitemShut {NoStop}%
\bibitem [{\citenamefont {Read}\ and\ \citenamefont
  {Green}(2000)}]{Read:2000zr}%
  \BibitemOpen
  \bibfield  {author} {\bibinfo {author} {\bibfnamefont {N.}~\bibnamefont
  {Read}}\ and\ \bibinfo {author} {\bibfnamefont {D.}~\bibnamefont {Green}},\
  }\href {http://arxiv.org/abs/cond-mat/9906453} {\bibfield  {journal}
  {\bibinfo  {journal} {Phys. Rev. B}\ }\textbf {\bibinfo {volume} {61}},\
  \bibinfo {pages} {10267} (\bibinfo {year} {2000})},\ \Eprint
  {http://arxiv.org/abs/cond-mat/9906453} {cond-mat/9906453} \BibitemShut
  {NoStop}%
\bibitem [{\citenamefont {Ivanov}(2001)}]{Ivanov01}%
  \BibitemOpen
  \bibfield  {author} {\bibinfo {author} {\bibfnamefont {D.~A.}\ \bibnamefont
  {Ivanov}},\ }\href {\doibase 10.1103/PhysRevLett.86.268} {\bibfield
  {journal} {\bibinfo  {journal} {Phys. Rev. Lett.}\ }\textbf {\bibinfo
  {volume} {86}},\ \bibinfo {pages} {268} (\bibinfo {year} {2001})}\BibitemShut
  {NoStop}%
\bibitem [{\citenamefont {Das~Sarma}\ \emph {et~al.}(2005)\citenamefont
  {Das~Sarma}, \citenamefont {Freedman},\ and\ \citenamefont
  {Nayak}}]{PhysRevLett.94.166802}%
  \BibitemOpen
  \bibfield  {author} {\bibinfo {author} {\bibfnamefont {S.}~\bibnamefont
  {Das~Sarma}}, \bibinfo {author} {\bibfnamefont {M.}~\bibnamefont {Freedman}},
  \ and\ \bibinfo {author} {\bibfnamefont {C.}~\bibnamefont {Nayak}},\ }\href
  {\doibase 10.1103/PhysRevLett.94.166802} {\bibfield  {journal} {\bibinfo
  {journal} {Phys. Rev. Lett.}\ }\textbf {\bibinfo {volume} {94}},\ \bibinfo
  {pages} {166802} (\bibinfo {year} {2005})}\BibitemShut {NoStop}%
\bibitem [{\citenamefont {Stern}\ and\ \citenamefont
  {Halperin}(2006)}]{PhysRevLett.96.016802}%
  \BibitemOpen
  \bibfield  {author} {\bibinfo {author} {\bibfnamefont {A.}~\bibnamefont
  {Stern}}\ and\ \bibinfo {author} {\bibfnamefont {B.~I.}\ \bibnamefont
  {Halperin}},\ }\href {\doibase 10.1103/PhysRevLett.96.016802} {\bibfield
  {journal} {\bibinfo  {journal} {Phys. Rev. Lett.}\ }\textbf {\bibinfo
  {volume} {96}},\ \bibinfo {pages} {016802} (\bibinfo {year}
  {2006})}\BibitemShut {NoStop}%
\bibitem [{\citenamefont {Bonderson}\ \emph {et~al.}(2006)\citenamefont
  {Bonderson}, \citenamefont {Kitaev},\ and\ \citenamefont
  {Shtengel}}]{PhysRevLett.96.016803}%
  \BibitemOpen
  \bibfield  {author} {\bibinfo {author} {\bibfnamefont {P.}~\bibnamefont
  {Bonderson}}, \bibinfo {author} {\bibfnamefont {A.}~\bibnamefont {Kitaev}}, \
  and\ \bibinfo {author} {\bibfnamefont {K.}~\bibnamefont {Shtengel}},\ }\href
  {\doibase 10.1103/PhysRevLett.96.016803} {\bibfield  {journal} {\bibinfo
  {journal} {Phys. Rev. Lett.}\ }\textbf {\bibinfo {volume} {96}},\ \bibinfo
  {pages} {016803} (\bibinfo {year} {2006})}\BibitemShut {NoStop}%
\bibitem [{\citenamefont {de~C.~Chamon}\ \emph {et~al.}(1997)\citenamefont
  {de~C.~Chamon}, \citenamefont {Freed}, \citenamefont {Kivelson},
  \citenamefont {Sondhi},\ and\ \citenamefont {Wen}}]{PhysRevB.55.2331}%
  \BibitemOpen
  \bibfield  {author} {\bibinfo {author} {\bibfnamefont {C.}~\bibnamefont
  {de~C.~Chamon}}, \bibinfo {author} {\bibfnamefont {D.~E.}\ \bibnamefont
  {Freed}}, \bibinfo {author} {\bibfnamefont {S.~A.}\ \bibnamefont {Kivelson}},
  \bibinfo {author} {\bibfnamefont {S.~L.}\ \bibnamefont {Sondhi}}, \ and\
  \bibinfo {author} {\bibfnamefont {X.~G.}\ \bibnamefont {Wen}},\ }\href
  {\doibase 10.1103/PhysRevB.55.2331} {\bibfield  {journal} {\bibinfo
  {journal} {Phys. Rev. B}\ }\textbf {\bibinfo {volume} {55}},\ \bibinfo
  {pages} {2331} (\bibinfo {year} {1997})}\BibitemShut {NoStop}%
\bibitem [{\citenamefont {Fradkin}\ \emph {et~al.}(1998)\citenamefont
  {Fradkin}, \citenamefont {Nayak}, \citenamefont {Tsvelik},\ and\
  \citenamefont {Wilczek}}]{Fradkin1998704}%
  \BibitemOpen
  \bibfield  {author} {\bibinfo {author} {\bibfnamefont {E.}~\bibnamefont
  {Fradkin}}, \bibinfo {author} {\bibfnamefont {C.}~\bibnamefont {Nayak}},
  \bibinfo {author} {\bibfnamefont {A.}~\bibnamefont {Tsvelik}}, \ and\
  \bibinfo {author} {\bibfnamefont {F.}~\bibnamefont {Wilczek}},\ }\href
  {\doibase http://dx.doi.org/10.1016/S0550-3213(98)00111-4} {\bibfield
  {journal} {\bibinfo  {journal} {Nuclear Physics B}\ }\textbf {\bibinfo
  {volume} {516}},\ \bibinfo {pages} {704 } (\bibinfo {year}
  {1998})}\BibitemShut {NoStop}%
\bibitem [{\citenamefont {Wen}(1995)}]{Wen:advances1995}%
  \BibitemOpen
  \bibfield  {author} {\bibinfo {author} {\bibfnamefont {X.-G.}\ \bibnamefont
  {Wen}},\ }\href {\doibase 10.1080/00018739500101566} {\bibfield  {journal}
  {\bibinfo  {journal} {Advances in Physics}\ }\textbf {\bibinfo {volume}
  {44}},\ \bibinfo {pages} {405} (\bibinfo {year} {1995})}\BibitemShut
  {NoStop}%
\bibitem [{\citenamefont {Chang}(2003)}]{Chang03}%
  \BibitemOpen
  \bibfield  {author} {\bibinfo {author} {\bibfnamefont {A.~M.}\ \bibnamefont
  {Chang}},\ }\href {\doibase 10.1103/RevModPhys.75.1449} {\bibfield  {journal}
  {\bibinfo  {journal} {Rev. Mod. Phys.}\ }\textbf {\bibinfo {volume} {75}},\
  \bibinfo {pages} {1449} (\bibinfo {year} {2003})}\BibitemShut {NoStop}%
\bibitem [{\citenamefont {Miller}\ \emph {et~al.}(2007)\citenamefont {Miller},
  \citenamefont {Radu}, \citenamefont {Zumb{\"u}hl}, \citenamefont
  {Levenson-Falk}, \citenamefont {Kastner}, \citenamefont {Marcus},
  \citenamefont {Pfeiffer},\ and\ \citenamefont {West}}]{Marcus:2007np}%
  \BibitemOpen
  \bibfield  {author} {\bibinfo {author} {\bibfnamefont {J.~B.}\ \bibnamefont
  {Miller}}, \bibinfo {author} {\bibfnamefont {I.~P.}\ \bibnamefont {Radu}},
  \bibinfo {author} {\bibfnamefont {D.~M.}\ \bibnamefont {Zumb{\"u}hl}},
  \bibinfo {author} {\bibfnamefont {E.~M.}\ \bibnamefont {Levenson-Falk}},
  \bibinfo {author} {\bibfnamefont {M.~A.}\ \bibnamefont {Kastner}}, \bibinfo
  {author} {\bibfnamefont {C.~M.}\ \bibnamefont {Marcus}}, \bibinfo {author}
  {\bibfnamefont {L.~N.}\ \bibnamefont {Pfeiffer}}, \ and\ \bibinfo {author}
  {\bibfnamefont {K.~W.}\ \bibnamefont {West}},\ }\href@noop {} {\bibfield
  {journal} {\bibinfo  {journal} {Nature Phys.}\ }\textbf {\bibinfo {volume}
  {3}},\ \bibinfo {pages} {561} (\bibinfo {year} {2007})}\BibitemShut {NoStop}%
\bibitem [{\citenamefont {Conti}\ and\ \citenamefont
  {Vignale}(1996)}]{Conti96}%
  \BibitemOpen
  \bibfield  {author} {\bibinfo {author} {\bibfnamefont {S.}~\bibnamefont
  {Conti}}\ and\ \bibinfo {author} {\bibfnamefont {G.}~\bibnamefont
  {Vignale}},\ }\href {\doibase 10.1103/PhysRevB.54.R14309} {\bibfield
  {journal} {\bibinfo  {journal} {Phys. Rev. B}\ }\textbf {\bibinfo {volume}
  {54}},\ \bibinfo {pages} {R14309} (\bibinfo {year} {1996})}\BibitemShut
  {NoStop}%
\bibitem [{\citenamefont {Shytov}\ \emph {et~al.}(1998)\citenamefont {Shytov},
  \citenamefont {Levitov},\ and\ \citenamefont {Halperin}}]{Shytov98}%
  \BibitemOpen
  \bibfield  {author} {\bibinfo {author} {\bibfnamefont {A.~V.}\ \bibnamefont
  {Shytov}}, \bibinfo {author} {\bibfnamefont {L.~S.}\ \bibnamefont {Levitov}},
  \ and\ \bibinfo {author} {\bibfnamefont {B.~I.}\ \bibnamefont {Halperin}},\
  }\href {\doibase 10.1103/PhysRevLett.80.141} {\bibfield  {journal} {\bibinfo
  {journal} {Phys. Rev. Lett.}\ }\textbf {\bibinfo {volume} {80}},\ \bibinfo
  {pages} {141} (\bibinfo {year} {1998})}\BibitemShut {NoStop}%
\bibitem [{\citenamefont {Lopez}\ and\ \citenamefont
  {Fradkin}(1999)}]{Lopez99}%
  \BibitemOpen
  \bibfield  {author} {\bibinfo {author} {\bibfnamefont {A.}~\bibnamefont
  {Lopez}}\ and\ \bibinfo {author} {\bibfnamefont {E.}~\bibnamefont
  {Fradkin}},\ }\href {\doibase 10.1103/PhysRevB.59.15323} {\bibfield
  {journal} {\bibinfo  {journal} {Phys. Rev. B}\ }\textbf {\bibinfo {volume}
  {59}},\ \bibinfo {pages} {15323} (\bibinfo {year} {1999})}\BibitemShut
  {NoStop}%
\bibitem [{\citenamefont {Mandal}\ and\ \citenamefont
  {Jain}(2001)}]{Mandal01b}%
  \BibitemOpen
  \bibfield  {author} {\bibinfo {author} {\bibfnamefont {S.~S.}\ \bibnamefont
  {Mandal}}\ and\ \bibinfo {author} {\bibfnamefont {J.}~\bibnamefont {Jain}},\
  }\href {\doibase http://dx.doi.org/10.1016/S0038-1098(01)00156-9} {\bibfield
  {journal} {\bibinfo  {journal} {Solid State Communications}\ }\textbf
  {\bibinfo {volume} {118}},\ \bibinfo {pages} {503 } (\bibinfo {year}
  {2001})}\BibitemShut {NoStop}%
\bibitem [{\citenamefont {Mandal}\ and\ \citenamefont
  {Jain}(2002)}]{Mandal02b}%
  \BibitemOpen
  \bibfield  {author} {\bibinfo {author} {\bibfnamefont {S.~S.}\ \bibnamefont
  {Mandal}}\ and\ \bibinfo {author} {\bibfnamefont {J.~K.}\ \bibnamefont
  {Jain}},\ }\href {\doibase 10.1103/PhysRevLett.89.096801} {\bibfield
  {journal} {\bibinfo  {journal} {Phys. Rev. Lett.}\ }\textbf {\bibinfo
  {volume} {89}},\ \bibinfo {pages} {096801} (\bibinfo {year}
  {2002})}\BibitemShut {NoStop}%
\bibitem [{\citenamefont {Z\"ulicke}\ \emph {et~al.}(2003)\citenamefont
  {Z\"ulicke}, \citenamefont {Palacios},\ and\ \citenamefont
  {MacDonald}}]{Zulicke03}%
  \BibitemOpen
  \bibfield  {author} {\bibinfo {author} {\bibfnamefont {U.}~\bibnamefont
  {Z\"ulicke}}, \bibinfo {author} {\bibfnamefont {J.~J.}\ \bibnamefont
  {Palacios}}, \ and\ \bibinfo {author} {\bibfnamefont {A.~H.}\ \bibnamefont
  {MacDonald}},\ }\href {\doibase 10.1103/PhysRevB.67.045303} {\bibfield
  {journal} {\bibinfo  {journal} {Phys. Rev. B}\ }\textbf {\bibinfo {volume}
  {67}},\ \bibinfo {pages} {045303} (\bibinfo {year} {2003})}\BibitemShut
  {NoStop}%
\bibitem [{\citenamefont {Wan}\ \emph {et~al.}(2005)\citenamefont {Wan},
  \citenamefont {Evers},\ and\ \citenamefont {Rezayi}}]{Wan05}%
  \BibitemOpen
  \bibfield  {author} {\bibinfo {author} {\bibfnamefont {X.}~\bibnamefont
  {Wan}}, \bibinfo {author} {\bibfnamefont {F.}~\bibnamefont {Evers}}, \ and\
  \bibinfo {author} {\bibfnamefont {E.~H.}\ \bibnamefont {Rezayi}},\ }\href
  {\doibase 10.1103/PhysRevLett.94.166804} {\bibfield  {journal} {\bibinfo
  {journal} {Phys. Rev. Lett.}\ }\textbf {\bibinfo {volume} {94}},\ \bibinfo
  {pages} {166804} (\bibinfo {year} {2005})}\BibitemShut {NoStop}%
\bibitem [{\citenamefont {Jolad}\ \emph {et~al.}(2007)\citenamefont {Jolad},
  \citenamefont {Chang},\ and\ \citenamefont {Jain}}]{Jolad07}%
  \BibitemOpen
  \bibfield  {author} {\bibinfo {author} {\bibfnamefont {S.}~\bibnamefont
  {Jolad}}, \bibinfo {author} {\bibfnamefont {C.-C.}\ \bibnamefont {Chang}}, \
  and\ \bibinfo {author} {\bibfnamefont {J.~K.}\ \bibnamefont {Jain}},\ }\href
  {\doibase 10.1103/PhysRevB.75.165306} {\bibfield  {journal} {\bibinfo
  {journal} {Phys. Rev. B}\ }\textbf {\bibinfo {volume} {75}},\ \bibinfo
  {pages} {165306} (\bibinfo {year} {2007})}\BibitemShut {NoStop}%
\bibitem [{\citenamefont {Jolad}\ and\ \citenamefont {Jain}(2009)}]{Jolad09}%
  \BibitemOpen
  \bibfield  {author} {\bibinfo {author} {\bibfnamefont {S.}~\bibnamefont
  {Jolad}}\ and\ \bibinfo {author} {\bibfnamefont {J.~K.}\ \bibnamefont
  {Jain}},\ }\href {\doibase 10.1103/PhysRevLett.102.116801} {\bibfield
  {journal} {\bibinfo  {journal} {Phys. Rev. Lett.}\ }\textbf {\bibinfo
  {volume} {102}},\ \bibinfo {pages} {116801} (\bibinfo {year}
  {2009})}\BibitemShut {NoStop}%
\bibitem [{\citenamefont {Wan}\ \emph {et~al.}(2002)\citenamefont {Wan},
  \citenamefont {Yang},\ and\ \citenamefont {Rezayi}}]{Wan02}%
  \BibitemOpen
  \bibfield  {author} {\bibinfo {author} {\bibfnamefont {X.}~\bibnamefont
  {Wan}}, \bibinfo {author} {\bibfnamefont {K.}~\bibnamefont {Yang}}, \ and\
  \bibinfo {author} {\bibfnamefont {E.~H.}\ \bibnamefont {Rezayi}},\ }\href
  {\doibase 10.1103/PhysRevLett.88.056802} {\bibfield  {journal} {\bibinfo
  {journal} {Phys. Rev. Lett.}\ }\textbf {\bibinfo {volume} {88}},\ \bibinfo
  {pages} {056802} (\bibinfo {year} {2002})}\BibitemShut {NoStop}%
\bibitem [{\citenamefont {Wan}\ \emph {et~al.}(2003)\citenamefont {Wan},
  \citenamefont {Rezayi},\ and\ \citenamefont {Yang}}]{Wan03}%
  \BibitemOpen
  \bibfield  {author} {\bibinfo {author} {\bibfnamefont {X.}~\bibnamefont
  {Wan}}, \bibinfo {author} {\bibfnamefont {E.~H.}\ \bibnamefont {Rezayi}}, \
  and\ \bibinfo {author} {\bibfnamefont {K.}~\bibnamefont {Yang}},\ }\href
  {\doibase 10.1103/PhysRevB.68.125307} {\bibfield  {journal} {\bibinfo
  {journal} {Phys. Rev. B}\ }\textbf {\bibinfo {volume} {68}},\ \bibinfo
  {pages} {125307} (\bibinfo {year} {2003})}\BibitemShut {NoStop}%
\bibitem [{\citenamefont {Chamon}\ and\ \citenamefont {Wen}(1994)}]{Chamon94}%
  \BibitemOpen
  \bibfield  {author} {\bibinfo {author} {\bibfnamefont {C.~d.~C.}\
  \bibnamefont {Chamon}}\ and\ \bibinfo {author} {\bibfnamefont {X.~G.}\
  \bibnamefont {Wen}},\ }\href {\doibase 10.1103/PhysRevB.49.8227} {\bibfield
  {journal} {\bibinfo  {journal} {Phys. Rev. B}\ }\textbf {\bibinfo {volume}
  {49}},\ \bibinfo {pages} {8227} (\bibinfo {year} {1994})}\BibitemShut
  {NoStop}%
\bibitem [{\citenamefont {Wan}\ \emph {et~al.}(2006)\citenamefont {Wan},
  \citenamefont {Yang},\ and\ \citenamefont {Rezayi}}]{Wan06}%
  \BibitemOpen
  \bibfield  {author} {\bibinfo {author} {\bibfnamefont {X.}~\bibnamefont
  {Wan}}, \bibinfo {author} {\bibfnamefont {K.}~\bibnamefont {Yang}}, \ and\
  \bibinfo {author} {\bibfnamefont {E.~H.}\ \bibnamefont {Rezayi}},\ }\href
  {\doibase 10.1103/PhysRevLett.97.256804} {\bibfield  {journal} {\bibinfo
  {journal} {Phys. Rev. Lett.}\ }\textbf {\bibinfo {volume} {97}},\ \bibinfo
  {pages} {256804} (\bibinfo {year} {2006})}\BibitemShut {NoStop}%
\bibitem [{\citenamefont {Wan}\ \emph {et~al.}(2008)\citenamefont {Wan},
  \citenamefont {Hu}, \citenamefont {Rezayi},\ and\ \citenamefont
  {Yang}}]{Wan08}%
  \BibitemOpen
  \bibfield  {author} {\bibinfo {author} {\bibfnamefont {X.}~\bibnamefont
  {Wan}}, \bibinfo {author} {\bibfnamefont {Z.-X.}\ \bibnamefont {Hu}},
  \bibinfo {author} {\bibfnamefont {E.~H.}\ \bibnamefont {Rezayi}}, \ and\
  \bibinfo {author} {\bibfnamefont {K.}~\bibnamefont {Yang}},\ }\href
  {http://arxiv.org/abs/0712.2095} {\bibfield  {journal} {\bibinfo  {journal}
  {Phys. Rev. B}\ }\textbf {\bibinfo {volume} {77}},\ \bibinfo {pages} {165316}
  (\bibinfo {year} {2008})},\ \Eprint {http://arxiv.org/abs/0712.2095}
  {0712.2095} \BibitemShut {NoStop}%
\bibitem [{\citenamefont {Wen}(1993)}]{Wen:1993}%
  \BibitemOpen
  \bibfield  {author} {\bibinfo {author} {\bibfnamefont {X.-G.}\ \bibnamefont
  {Wen}},\ }\href@noop {} {\bibfield  {journal} {\bibinfo  {journal} {Phys.
  Rev. Lett.}\ }\textbf {\bibinfo {volume} {70}},\ \bibinfo {pages} {355}
  (\bibinfo {year} {1993})}\BibitemShut {NoStop}%
\bibitem [{\citenamefont {Read}\ and\ \citenamefont
  {Rezayi}(1999)}]{PhysRevB.59.8084}%
  \BibitemOpen
  \bibfield  {author} {\bibinfo {author} {\bibfnamefont {N.}~\bibnamefont
  {Read}}\ and\ \bibinfo {author} {\bibfnamefont {E.}~\bibnamefont {Rezayi}},\
  }\href {\doibase 10.1103/PhysRevB.59.8084} {\bibfield  {journal} {\bibinfo
  {journal} {Phys. Rev. B}\ }\textbf {\bibinfo {volume} {59}},\ \bibinfo
  {pages} {8084} (\bibinfo {year} {1999})}\BibitemShut {NoStop}%
\bibitem [{\citenamefont {Lee}\ \emph {et~al.}(2014)\citenamefont {Lee},
  \citenamefont {Hu},\ and\ \citenamefont {Wan}}]{Lee14}%
  \BibitemOpen
  \bibfield  {author} {\bibinfo {author} {\bibfnamefont {K.~H.}\ \bibnamefont
  {Lee}}, \bibinfo {author} {\bibfnamefont {Z.-X.}\ \bibnamefont {Hu}}, \ and\
  \bibinfo {author} {\bibfnamefont {X.}~\bibnamefont {Wan}},\ }\href {\doibase
  10.1103/PhysRevB.89.165124} {\bibfield  {journal} {\bibinfo  {journal} {Phys.
  Rev. B}\ }\textbf {\bibinfo {volume} {89}},\ \bibinfo {pages} {165124}
  (\bibinfo {year} {2014})}\BibitemShut {NoStop}%
\bibitem [{\citenamefont {Baer}\ \emph {et~al.}(2014)\citenamefont {Baer},
  \citenamefont {R\"ossler}, \citenamefont {Ihn}, \citenamefont {Ensslin},
  \citenamefont {Reichl},\ and\ \citenamefont {Wegscheider}}]{Baer14}%
  \BibitemOpen
  \bibfield  {author} {\bibinfo {author} {\bibfnamefont {S.}~\bibnamefont
  {Baer}}, \bibinfo {author} {\bibfnamefont {C.}~\bibnamefont {R\"ossler}},
  \bibinfo {author} {\bibfnamefont {T.}~\bibnamefont {Ihn}}, \bibinfo {author}
  {\bibfnamefont {K.}~\bibnamefont {Ensslin}}, \bibinfo {author} {\bibfnamefont
  {C.}~\bibnamefont {Reichl}}, \ and\ \bibinfo {author} {\bibfnamefont
  {W.}~\bibnamefont {Wegscheider}},\ }\href {\doibase
  10.1103/PhysRevB.90.075403} {\bibfield  {journal} {\bibinfo  {journal} {Phys.
  Rev. B}\ }\textbf {\bibinfo {volume} {90}},\ \bibinfo {pages} {075403}
  (\bibinfo {year} {2014})}\BibitemShut {NoStop}%
\bibitem [{\citenamefont {Dolev}\ \emph {et~al.}(2011)\citenamefont {Dolev},
  \citenamefont {Gross}, \citenamefont {Sabo}, \citenamefont {Gurman},
  \citenamefont {M.~Heiblum},\ and\ \citenamefont {Mahalu}}]{73exp1}%
  \BibitemOpen
  \bibfield  {author} {\bibinfo {author} {\bibfnamefont {M.}~\bibnamefont
  {Dolev}}, \bibinfo {author} {\bibfnamefont {Y.}~\bibnamefont {Gross}},
  \bibinfo {author} {\bibfnamefont {R.}~\bibnamefont {Sabo}}, \bibinfo {author}
  {\bibfnamefont {I.}~\bibnamefont {Gurman}}, \bibinfo {author} {\bibfnamefont
  {V.~U.}\ \bibnamefont {M.~Heiblum}}, \ and\ \bibinfo {author} {\bibfnamefont
  {D.}~\bibnamefont {Mahalu}},\ }\href@noop {} {\bibfield  {journal} {\bibinfo
  {journal} {Phys. Rev. Lett.}\ }\textbf {\bibinfo {volume} {107}},\ \bibinfo
  {pages} {036805} (\bibinfo {year} {2011})}\BibitemShut {NoStop}%
\bibitem [{\citenamefont {Gross}\ \emph {et~al.}(2012)\citenamefont {Gross},
  \citenamefont {Dolev}, \citenamefont {Heiblum}, \citenamefont {Umansky},\
  and\ \citenamefont {Mahalu}}]{73exp2}%
  \BibitemOpen
  \bibfield  {author} {\bibinfo {author} {\bibfnamefont {Y.}~\bibnamefont
  {Gross}}, \bibinfo {author} {\bibfnamefont {M.}~\bibnamefont {Dolev}},
  \bibinfo {author} {\bibfnamefont {M.}~\bibnamefont {Heiblum}}, \bibinfo
  {author} {\bibfnamefont {V.}~\bibnamefont {Umansky}}, \ and\ \bibinfo
  {author} {\bibfnamefont {D.}~\bibnamefont {Mahalu}},\ }\href {\doibase
  10.1103/PhysRevLett.108.226801} {\bibfield  {journal} {\bibinfo  {journal}
  {Phys. Rev. Lett.}\ }\textbf {\bibinfo {volume} {108}},\ \bibinfo {pages}
  {226801} (\bibinfo {year} {2012})}\BibitemShut {NoStop}%
\bibitem [{\citenamefont {Venkatachalam}\ \emph {et~al.}(2011)\citenamefont
  {Venkatachalam}, \citenamefont {Yacoby}, \citenamefont {Pfeiffer},\ and\
  \citenamefont {West}}]{73exp3}%
  \BibitemOpen
  \bibfield  {author} {\bibinfo {author} {\bibfnamefont {V.}~\bibnamefont
  {Venkatachalam}}, \bibinfo {author} {\bibfnamefont {A.}~\bibnamefont
  {Yacoby}}, \bibinfo {author} {\bibfnamefont {L.}~\bibnamefont {Pfeiffer}}, \
  and\ \bibinfo {author} {\bibfnamefont {K.}~\bibnamefont {West}},\ }\href@noop
  {} {\bibfield  {journal} {\bibinfo  {journal} {Nature}\ }\textbf {\bibinfo
  {volume} {469}},\ \bibinfo {pages} {185} (\bibinfo {year}
  {2011})}\BibitemShut {NoStop}%
\bibitem [{\citenamefont {Willett}\ \emph {et~al.}(2009)\citenamefont
  {Willett}, \citenamefont {Pfeiffer},\ and\ \citenamefont {West}}]{73exp4}%
  \BibitemOpen
  \bibfield  {author} {\bibinfo {author} {\bibfnamefont {R.~L.}\ \bibnamefont
  {Willett}}, \bibinfo {author} {\bibfnamefont {L.~N.}\ \bibnamefont
  {Pfeiffer}}, \ and\ \bibinfo {author} {\bibfnamefont {K.~W.}\ \bibnamefont
  {West}},\ }\href@noop {} {\bibfield  {journal} {\bibinfo  {journal} {Proc.
  Nat. Acad. Sci. (US)}\ }\textbf {\bibinfo {volume} {106}},\ \bibinfo {pages}
  {8853} (\bibinfo {year} {2009})}\BibitemShut {NoStop}%
\bibitem [{\citenamefont {An}\ \emph {et~al.}(2011)\citenamefont {An},
  \citenamefont {Jiang}, \citenamefont {Choi}, \citenamefont {Kang},
  \citenamefont {Simon}, \citenamefont {Pfeiffer}, \citenamefont {West},\ and\
  \citenamefont {Baldwin}}]{73exp5}%
  \BibitemOpen
  \bibfield  {author} {\bibinfo {author} {\bibfnamefont {S.}~\bibnamefont
  {An}}, \bibinfo {author} {\bibfnamefont {P.}~\bibnamefont {Jiang}}, \bibinfo
  {author} {\bibfnamefont {H.}~\bibnamefont {Choi}}, \bibinfo {author}
  {\bibfnamefont {W.}~\bibnamefont {Kang}}, \bibinfo {author} {\bibfnamefont
  {S.~H.}\ \bibnamefont {Simon}}, \bibinfo {author} {\bibfnamefont {L.~N.}\
  \bibnamefont {Pfeiffer}}, \bibinfo {author} {\bibfnamefont {K.~W.}\
  \bibnamefont {West}}, \ and\ \bibinfo {author} {\bibfnamefont {K.~W.}\
  \bibnamefont {Baldwin}},\ }\href {http://arxiv.org/abs/1112.3400} {\bibfield
  {journal} {\bibinfo  {journal} {ArXiv e-prints}\ } (\bibinfo {year}
  {2011})},\ \Eprint {http://arxiv.org/abs/1112.3400} {arXiv:1112.3400
  ["cond-mat.mes-hall"]} \BibitemShut {NoStop}%
\bibitem [{\citenamefont {Xia}\ \emph {et~al.}(2011)\citenamefont {Xia},
  \citenamefont {Eisenstein}, \citenamefont {Pfeiffer},\ and\ \citenamefont
  {West}}]{73exp6}%
  \BibitemOpen
  \bibfield  {author} {\bibinfo {author} {\bibfnamefont {J.}~\bibnamefont
  {Xia}}, \bibinfo {author} {\bibfnamefont {J.~P.}\ \bibnamefont {Eisenstein}},
  \bibinfo {author} {\bibfnamefont {L.~N.}\ \bibnamefont {Pfeiffer}}, \ and\
  \bibinfo {author} {\bibfnamefont {K.~W.}\ \bibnamefont {West}},\ }\href@noop
  {} {\bibfield  {journal} {\bibinfo  {journal} {Nature Phys.}\ }\textbf
  {\bibinfo {volume} {7}},\ \bibinfo {pages} {845} (\bibinfo {year}
  {2011})}\BibitemShut {NoStop}%
\bibitem [{\citenamefont {Wurstbauer}\ \emph {et~al.}(2013)\citenamefont
  {Wurstbauer}, \citenamefont {West}, \citenamefont {Pfeiffer},\ and\
  \citenamefont {Pinczuk}}]{73exp7}%
  \BibitemOpen
  \bibfield  {author} {\bibinfo {author} {\bibfnamefont {U.}~\bibnamefont
  {Wurstbauer}}, \bibinfo {author} {\bibfnamefont {K.~W.}\ \bibnamefont
  {West}}, \bibinfo {author} {\bibfnamefont {L.~N.}\ \bibnamefont {Pfeiffer}},
  \ and\ \bibinfo {author} {\bibfnamefont {A.}~\bibnamefont {Pinczuk}},\
  }\href@noop {} {\bibfield  {journal} {\bibinfo  {journal} {Phys. Rev. Lett.}\
  }\textbf {\bibinfo {volume} {110}},\ \bibinfo {pages} {026801} (\bibinfo
  {year} {2013})}\BibitemShut {NoStop}%
\bibitem [{\citenamefont {Balram}\ \emph
  {et~al.}(2013{\natexlab{a}})\citenamefont {Balram}, \citenamefont {W\'ojs},\
  and\ \citenamefont {Jain}}]{Balram13}%
  \BibitemOpen
  \bibfield  {author} {\bibinfo {author} {\bibfnamefont {A.~C.}\ \bibnamefont
  {Balram}}, \bibinfo {author} {\bibfnamefont {A.}~\bibnamefont {W\'ojs}}, \
  and\ \bibinfo {author} {\bibfnamefont {J.~K.}\ \bibnamefont {Jain}},\ }\href
  {\doibase 10.1103/PhysRevB.88.205312} {\bibfield  {journal} {\bibinfo
  {journal} {Phys. Rev. B}\ }\textbf {\bibinfo {volume} {88}},\ \bibinfo
  {pages} {205312} (\bibinfo {year} {2013}{\natexlab{a}})}\BibitemShut
  {NoStop}%
\bibitem [{\citenamefont {Halperin}(1983)}]{Halperin83}%
  \BibitemOpen
  \bibfield  {author} {\bibinfo {author} {\bibfnamefont {B.~I.}\ \bibnamefont
  {Halperin}},\ }\href@noop {} {\bibfield  {journal} {\bibinfo  {journal}
  {Helv. Phys. Acta}\ }\textbf {\bibinfo {volume} {56}},\ \bibinfo {pages} {75}
  (\bibinfo {year} {1983})}\BibitemShut {NoStop}%
\bibitem [{\citenamefont {Lin}\ \emph {et~al.}(2012)\citenamefont {Lin},
  \citenamefont {Dillard}, \citenamefont {Kastner}, \citenamefont {Pfeiffer},\
  and\ \citenamefont {West}}]{Lin12}%
  \BibitemOpen
  \bibfield  {author} {\bibinfo {author} {\bibfnamefont {X.}~\bibnamefont
  {Lin}}, \bibinfo {author} {\bibfnamefont {C.}~\bibnamefont {Dillard}},
  \bibinfo {author} {\bibfnamefont {M.~A.}\ \bibnamefont {Kastner}}, \bibinfo
  {author} {\bibfnamefont {L.~N.}\ \bibnamefont {Pfeiffer}}, \ and\ \bibinfo
  {author} {\bibfnamefont {K.~W.}\ \bibnamefont {West}},\ }\href {\doibase
  10.1103/PhysRevB.85.165321} {\bibfield  {journal} {\bibinfo  {journal} {Phys.
  Rev. B}\ }\textbf {\bibinfo {volume} {85}},\ \bibinfo {pages} {165321}
  (\bibinfo {year} {2012})}\BibitemShut {NoStop}%
\bibitem [{\citenamefont {Yang}\ and\ \citenamefont {Feldman}(2014)}]{Yang14b}%
  \BibitemOpen
  \bibfield  {author} {\bibinfo {author} {\bibfnamefont {G.}~\bibnamefont
  {Yang}}\ and\ \bibinfo {author} {\bibfnamefont {D.~E.}\ \bibnamefont
  {Feldman}},\ }\href@noop {} {\bibfield  {journal} {\bibinfo  {journal} {ArXiv
  e-prints}\ } (\bibinfo {year} {2014})},\ \Eprint
  {http://arxiv.org/abs/1406.2263} {arXiv:1406.2263 ["cond-mat.mes-hall"]}
  \BibitemShut {NoStop}%
\bibitem [{\citenamefont {Balram}\ \emph
  {et~al.}(2013{\natexlab{b}})\citenamefont {Balram}, \citenamefont {Wu},
  \citenamefont {Sreejith}, \citenamefont {W\'ojs},\ and\ \citenamefont
  {Jain}}]{Balram13b}%
  \BibitemOpen
  \bibfield  {author} {\bibinfo {author} {\bibfnamefont {A.~C.}\ \bibnamefont
  {Balram}}, \bibinfo {author} {\bibfnamefont {Y.-H.}\ \bibnamefont {Wu}},
  \bibinfo {author} {\bibfnamefont {G.~J.}\ \bibnamefont {Sreejith}}, \bibinfo
  {author} {\bibfnamefont {A.}~\bibnamefont {W\'ojs}}, \ and\ \bibinfo {author}
  {\bibfnamefont {J.~K.}\ \bibnamefont {Jain}},\ }\href {\doibase
  10.1103/PhysRevLett.110.186801} {\bibfield  {journal} {\bibinfo  {journal}
  {Phys. Rev. Lett.}\ }\textbf {\bibinfo {volume} {110}},\ \bibinfo {pages}
  {186801} (\bibinfo {year} {2013}{\natexlab{b}})}\BibitemShut {NoStop}%
\bibitem [{\citenamefont {Chklovskii}(1995)}]{PhysRevB.51.9895}%
  \BibitemOpen
  \bibfield  {author} {\bibinfo {author} {\bibfnamefont {D.~B.}\ \bibnamefont
  {Chklovskii}},\ }\href {\doibase 10.1103/PhysRevB.51.9895} {\bibfield
  {journal} {\bibinfo  {journal} {Phys. Rev. B}\ }\textbf {\bibinfo {volume}
  {51}},\ \bibinfo {pages} {9895} (\bibinfo {year} {1995})}\BibitemShut
  {NoStop}%
\bibitem [{\citenamefont {Storni}\ \emph {et~al.}(2010)\citenamefont {Storni},
  \citenamefont {Morf},\ and\ \citenamefont
  {Das~Sarma}}]{PhysRevLett.104.076803}%
  \BibitemOpen
  \bibfield  {author} {\bibinfo {author} {\bibfnamefont {M.}~\bibnamefont
  {Storni}}, \bibinfo {author} {\bibfnamefont {R.~H.}\ \bibnamefont {Morf}}, \
  and\ \bibinfo {author} {\bibfnamefont {S.}~\bibnamefont {Das~Sarma}},\ }\href
  {\doibase 10.1103/PhysRevLett.104.076803} {\bibfield  {journal} {\bibinfo
  {journal} {Phys. Rev. Lett.}\ }\textbf {\bibinfo {volume} {104}},\ \bibinfo
  {pages} {076803} (\bibinfo {year} {2010})}\BibitemShut {NoStop}%
\bibitem [{\citenamefont {Park}\ \emph {et~al.}(1998)\citenamefont {Park},
  \citenamefont {Melik-Alaverdian}, \citenamefont {Bonesteel},\ and\
  \citenamefont {Jain}}]{PhysRevB.58.R10167}%
  \BibitemOpen
  \bibfield  {author} {\bibinfo {author} {\bibfnamefont {K.}~\bibnamefont
  {Park}}, \bibinfo {author} {\bibfnamefont {V.}~\bibnamefont
  {Melik-Alaverdian}}, \bibinfo {author} {\bibfnamefont {N.~E.}\ \bibnamefont
  {Bonesteel}}, \ and\ \bibinfo {author} {\bibfnamefont {J.~K.}\ \bibnamefont
  {Jain}},\ }\href {\doibase 10.1103/PhysRevB.58.R10167} {\bibfield  {journal}
  {\bibinfo  {journal} {Phys. Rev. B}\ }\textbf {\bibinfo {volume} {58}},\
  \bibinfo {pages} {R10167} (\bibinfo {year} {1998})}\BibitemShut {NoStop}%
\bibitem [{\citenamefont {Bernevig}\ and\ \citenamefont
  {Haldane}(2008)}]{Bernevig08}%
  \BibitemOpen
  \bibfield  {author} {\bibinfo {author} {\bibfnamefont {B.~A.}\ \bibnamefont
  {Bernevig}}\ and\ \bibinfo {author} {\bibfnamefont {F.~D.~M.}\ \bibnamefont
  {Haldane}},\ }\href {\doibase 10.1103/PhysRevLett.100.246802} {\bibfield
  {journal} {\bibinfo  {journal} {Phys. Rev. Lett.}\ }\textbf {\bibinfo
  {volume} {100}},\ \bibinfo {pages} {246802} (\bibinfo {year}
  {2008})}\BibitemShut {NoStop}%
\bibitem [{\citenamefont {Bernevig}\ and\ \citenamefont
  {Regnault}(2009)}]{Bernevig09}%
  \BibitemOpen
  \bibfield  {author} {\bibinfo {author} {\bibfnamefont {B.~A.}\ \bibnamefont
  {Bernevig}}\ and\ \bibinfo {author} {\bibfnamefont {N.}~\bibnamefont
  {Regnault}},\ }\href {\doibase 10.1103/PhysRevLett.103.206801} {\bibfield
  {journal} {\bibinfo  {journal} {Phys. Rev. Lett.}\ }\textbf {\bibinfo
  {volume} {103}},\ \bibinfo {pages} {206801} (\bibinfo {year}
  {2009})}\BibitemShut {NoStop}%
\bibitem [{\citenamefont {Thomale}\ \emph {et~al.}(2011)\citenamefont
  {Thomale}, \citenamefont {Estienne}, \citenamefont {Regnault},\ and\
  \citenamefont {Bernevig}}]{Thomale11}%
  \BibitemOpen
  \bibfield  {author} {\bibinfo {author} {\bibfnamefont {R.}~\bibnamefont
  {Thomale}}, \bibinfo {author} {\bibfnamefont {B.}~\bibnamefont {Estienne}},
  \bibinfo {author} {\bibfnamefont {N.}~\bibnamefont {Regnault}}, \ and\
  \bibinfo {author} {\bibfnamefont {B.~A.}\ \bibnamefont {Bernevig}},\ }\href
  {\doibase 10.1103/PhysRevB.84.045127} {\bibfield  {journal} {\bibinfo
  {journal} {Phys. Rev. B}\ }\textbf {\bibinfo {volume} {84}},\ \bibinfo
  {pages} {045127} (\bibinfo {year} {2011})}\BibitemShut {NoStop}%
\bibitem [{\citenamefont {Oaknin}\ \emph {et~al.}(1995)\citenamefont {Oaknin},
  \citenamefont {Martin-Moreno}, \citenamefont {Palacios},\ and\ \citenamefont
  {Tejedor}}]{PhysRevLett.74.5120}%
  \BibitemOpen
  \bibfield  {author} {\bibinfo {author} {\bibfnamefont {J.~H.}\ \bibnamefont
  {Oaknin}}, \bibinfo {author} {\bibfnamefont {L.}~\bibnamefont
  {Martin-Moreno}}, \bibinfo {author} {\bibfnamefont {J.~J.}\ \bibnamefont
  {Palacios}}, \ and\ \bibinfo {author} {\bibfnamefont {C.}~\bibnamefont
  {Tejedor}},\ }\href {\doibase 10.1103/PhysRevLett.74.5120} {\bibfield
  {journal} {\bibinfo  {journal} {Phys. Rev. Lett.}\ }\textbf {\bibinfo
  {volume} {74}},\ \bibinfo {pages} {5120} (\bibinfo {year}
  {1995})}\BibitemShut {NoStop}%
\bibitem [{\citenamefont {Jolad}\ \emph {et~al.}(2010)\citenamefont {Jolad},
  \citenamefont {Sen},\ and\ \citenamefont {Jain}}]{Jolad:2010}%
  \BibitemOpen
  \bibfield  {author} {\bibinfo {author} {\bibfnamefont {S.}~\bibnamefont
  {Jolad}}, \bibinfo {author} {\bibfnamefont {D.}~\bibnamefont {Sen}}, \ and\
  \bibinfo {author} {\bibfnamefont {J.~K.}\ \bibnamefont {Jain}},\ }\href
  {\doibase 10.1103/PhysRevB.82.075315} {\bibfield  {journal} {\bibinfo
  {journal} {Phys. Rev. B}\ }\textbf {\bibinfo {volume} {82}},\ \bibinfo
  {pages} {075315} (\bibinfo {year} {2010})}\BibitemShut {NoStop}%
\bibitem [{\citenamefont {Milovanovi\ifmmode~\acute{c}\else \'{c}\fi{}}\ and\
  \citenamefont {Read}(1996)}]{PhysRevB.53.13559}%
  \BibitemOpen
  \bibfield  {author} {\bibinfo {author} {\bibfnamefont {M.}~\bibnamefont
  {Milovanovi\ifmmode~\acute{c}\else \'{c}\fi{}}}\ and\ \bibinfo {author}
  {\bibfnamefont {N.}~\bibnamefont {Read}},\ }\href {\doibase
  10.1103/PhysRevB.53.13559} {\bibfield  {journal} {\bibinfo  {journal} {Phys.
  Rev. B}\ }\textbf {\bibinfo {volume} {53}},\ \bibinfo {pages} {13559}
  (\bibinfo {year} {1996})}\BibitemShut {NoStop}%
\bibitem [{\citenamefont {Metropolis}\ \emph {et~al.}(1953)\citenamefont
  {Metropolis}, \citenamefont {Rosenbluth}, \citenamefont {Rosenbluth},
  \citenamefont {Teller},\ and\ \citenamefont {Teller}}]{Metropolis:1953}%
  \BibitemOpen
  \bibfield  {author} {\bibinfo {author} {\bibfnamefont {N.}~\bibnamefont
  {Metropolis}}, \bibinfo {author} {\bibfnamefont {A.~W.}\ \bibnamefont
  {Rosenbluth}}, \bibinfo {author} {\bibfnamefont {M.~N.}\ \bibnamefont
  {Rosenbluth}}, \bibinfo {author} {\bibfnamefont {A.~H.}\ \bibnamefont
  {Teller}}, \ and\ \bibinfo {author} {\bibfnamefont {E.}~\bibnamefont
  {Teller}},\ }\href@noop {} {\bibfield  {journal} {\bibinfo  {journal} {J.
  Chem. Phys.}\ }\textbf {\bibinfo {volume} {21}},\ \bibinfo {pages} {1087}
  (\bibinfo {year} {1953})}\BibitemShut {NoStop}%
\bibitem [{\citenamefont {Jain}(2007)}]{Jain07}%
  \BibitemOpen
  \bibfield  {author} {\bibinfo {author} {\bibfnamefont {J.~K.}\ \bibnamefont
  {Jain}},\ }\href@noop {} {\emph {\bibinfo {title} {Composite Fermions}}}\
  (\bibinfo  {publisher} {Cambridge University Press, New York, US},\ \bibinfo
  {year} {2007})\BibitemShut {NoStop}%
\bibitem [{\citenamefont {Shi}\ \emph {et~al.}(2008)\citenamefont {Shi},
  \citenamefont {Jolad}, \citenamefont {Regnault},\ and\ \citenamefont
  {Jain}}]{ChuntaiShi:2008}%
  \BibitemOpen
  \bibfield  {author} {\bibinfo {author} {\bibfnamefont {C.}~\bibnamefont
  {Shi}}, \bibinfo {author} {\bibfnamefont {S.}~\bibnamefont {Jolad}}, \bibinfo
  {author} {\bibfnamefont {N.}~\bibnamefont {Regnault}}, \ and\ \bibinfo
  {author} {\bibfnamefont {J.~K.}\ \bibnamefont {Jain}},\ }\href {\doibase
  10.1103/PhysRevB.77.155127} {\bibfield  {journal} {\bibinfo  {journal} {Phys.
  Rev. B}\ }\textbf {\bibinfo {volume} {77}},\ \bibinfo {pages} {155127}
  (\bibinfo {year} {2008})}\BibitemShut {NoStop}%
\bibitem [{\citenamefont {Zhang}\ \emph {et~al.}(2013)\citenamefont {Zhang},
  \citenamefont {Hu},\ and\ \citenamefont {Yang}}]{Zhang13}%
  \BibitemOpen
  \bibfield  {author} {\bibinfo {author} {\bibfnamefont {Y.}~\bibnamefont
  {Zhang}}, \bibinfo {author} {\bibfnamefont {Z.-X.}\ \bibnamefont {Hu}}, \
  and\ \bibinfo {author} {\bibfnamefont {K.}~\bibnamefont {Yang}},\ }\href
  {\doibase 10.1103/PhysRevB.88.205128} {\bibfield  {journal} {\bibinfo
  {journal} {Phys. Rev. B}\ }\textbf {\bibinfo {volume} {88}},\ \bibinfo
  {pages} {205128} (\bibinfo {year} {2013})}\BibitemShut {NoStop}%
\bibitem [{\citenamefont {Yang}(2003)}]{PhysRevLett.91.036802}%
  \BibitemOpen
  \bibfield  {author} {\bibinfo {author} {\bibfnamefont {K.}~\bibnamefont
  {Yang}},\ }\href {\doibase 10.1103/PhysRevLett.91.036802} {\bibfield
  {journal} {\bibinfo  {journal} {Phys. Rev. Lett.}\ }\textbf {\bibinfo
  {volume} {91}},\ \bibinfo {pages} {036802} (\bibinfo {year}
  {2003})}\BibitemShut {NoStop}%
\bibitem [{\citenamefont {Moore}\ and\ \citenamefont
  {Wen}(1998)}]{PhysRevB.57.10138}%
  \BibitemOpen
  \bibfield  {author} {\bibinfo {author} {\bibfnamefont {J.~E.}\ \bibnamefont
  {Moore}}\ and\ \bibinfo {author} {\bibfnamefont {X.-G.}\ \bibnamefont
  {Wen}},\ }\href {\doibase 10.1103/PhysRevB.57.10138} {\bibfield  {journal}
  {\bibinfo  {journal} {Phys. Rev. B}\ }\textbf {\bibinfo {volume} {57}},\
  \bibinfo {pages} {10138} (\bibinfo {year} {1998})}\BibitemShut {NoStop}%
\bibitem [{\citenamefont {Overbosch}\ and\ \citenamefont
  {Wen}(2008)}]{overbosch2008phase}%
  \BibitemOpen
  \bibfield  {author} {\bibinfo {author} {\bibfnamefont {B.}~\bibnamefont
  {Overbosch}}\ and\ \bibinfo {author} {\bibfnamefont {X.-G.}\ \bibnamefont
  {Wen}},\ }\href@noop {} {\bibfield  {journal} {\bibinfo  {journal} {arXiv
  preprint arXiv:0804.2087}\ } (\bibinfo {year} {2008})}\BibitemShut {NoStop}%
\bibitem [{\citenamefont {Kane}\ \emph {et~al.}(1994)\citenamefont {Kane},
  \citenamefont {Fisher},\ and\ \citenamefont
  {Polchinski}}]{PhysRevLett.72.4129}%
  \BibitemOpen
  \bibfield  {author} {\bibinfo {author} {\bibfnamefont {C.~L.}\ \bibnamefont
  {Kane}}, \bibinfo {author} {\bibfnamefont {M.~P.~A.}\ \bibnamefont {Fisher}},
  \ and\ \bibinfo {author} {\bibfnamefont {J.}~\bibnamefont {Polchinski}},\
  }\href {\doibase 10.1103/PhysRevLett.72.4129} {\bibfield  {journal} {\bibinfo
   {journal} {Phys. Rev. Lett.}\ }\textbf {\bibinfo {volume} {72}},\ \bibinfo
  {pages} {4129} (\bibinfo {year} {1994})}\BibitemShut {NoStop}%
\bibitem [{\citenamefont {Bid}\ \emph {et~al.}(2010)\citenamefont {Bid},
  \citenamefont {Ofek}, \citenamefont {Inoue}, \citenamefont {Heiblum},
  \citenamefont {Kane}, \citenamefont {Umansky},\ and\ \citenamefont
  {Mahalu}}]{bid2010observation}%
  \BibitemOpen
  \bibfield  {author} {\bibinfo {author} {\bibfnamefont {A.}~\bibnamefont
  {Bid}}, \bibinfo {author} {\bibfnamefont {N.}~\bibnamefont {Ofek}}, \bibinfo
  {author} {\bibfnamefont {H.}~\bibnamefont {Inoue}}, \bibinfo {author}
  {\bibfnamefont {M.}~\bibnamefont {Heiblum}}, \bibinfo {author} {\bibfnamefont
  {C.}~\bibnamefont {Kane}}, \bibinfo {author} {\bibfnamefont {V.}~\bibnamefont
  {Umansky}}, \ and\ \bibinfo {author} {\bibfnamefont {D.}~\bibnamefont
  {Mahalu}},\ }\href@noop {} {\bibfield  {journal} {\bibinfo  {journal}
  {Nature}\ }\textbf {\bibinfo {volume} {466}},\ \bibinfo {pages} {585}
  (\bibinfo {year} {2010})}\BibitemShut {NoStop}%
\bibitem [{\citenamefont {Carrega}\ \emph {et~al.}(2011)\citenamefont
  {Carrega}, \citenamefont {Ferraro}, \citenamefont {Braggio}, \citenamefont
  {Magnoli},\ and\ \citenamefont {Sassetti}}]{Carrega11}%
  \BibitemOpen
  \bibfield  {author} {\bibinfo {author} {\bibfnamefont {M.}~\bibnamefont
  {Carrega}}, \bibinfo {author} {\bibfnamefont {D.}~\bibnamefont {Ferraro}},
  \bibinfo {author} {\bibfnamefont {A.}~\bibnamefont {Braggio}}, \bibinfo
  {author} {\bibfnamefont {N.}~\bibnamefont {Magnoli}}, \ and\ \bibinfo
  {author} {\bibfnamefont {M.}~\bibnamefont {Sassetti}},\ }\href {\doibase
  10.1103/PhysRevLett.107.146404} {\bibfield  {journal} {\bibinfo  {journal}
  {Phys. Rev. Lett.}\ }\textbf {\bibinfo {volume} {107}},\ \bibinfo {pages}
  {146404} (\bibinfo {year} {2011})}\BibitemShut {NoStop}%
\bibitem [{\citenamefont {Carrega}\ \emph {et~al.}(2012)\citenamefont
  {Carrega}, \citenamefont {Ferraro}, \citenamefont {Braggio}, \citenamefont
  {Magnoli},\ and\ \citenamefont {Sassetti}}]{Carrega12}%
  \BibitemOpen
  \bibfield  {author} {\bibinfo {author} {\bibfnamefont {M.}~\bibnamefont
  {Carrega}}, \bibinfo {author} {\bibfnamefont {D.}~\bibnamefont {Ferraro}},
  \bibinfo {author} {\bibfnamefont {A.}~\bibnamefont {Braggio}}, \bibinfo
  {author} {\bibfnamefont {N.}~\bibnamefont {Magnoli}}, \ and\ \bibinfo
  {author} {\bibfnamefont {M.}~\bibnamefont {Sassetti}},\ }\href
  {http://stacks.iop.org/1367-2630/14/i=2/a=023017} {\bibfield  {journal}
  {\bibinfo  {journal} {New Journal of Physics}\ }\textbf {\bibinfo {volume}
  {14}},\ \bibinfo {pages} {023017} (\bibinfo {year} {2012})}\BibitemShut
  {NoStop}%
\end{thebibliography}%
\end{document}